%% file: CodeComparison.tex
\documentclass[twoside]{article}
\usepackage{qic,epsfig}

\usepackage{cite}
\usepackage{booktabs}
\newcommand{\ra}[1]{\renewcommand{\arraystretch}{#1}}
\usepackage{multirow}
\usepackage[cmex10]{amsmath}
\usepackage{braket}
\usepackage{bm}
\usepackage{graphicx}
\usepackage{rotating}
\usepackage{enumerate}
\usepackage{subfigure}
\usepackage[pdftex]{color}
\newcommand{\ignore}[1]{}

\begin{document}
\setlength{\textheight}{8.0truein}    

\runninghead{Comparing the Overhead of Topological and Concatenated Quantum Error Correction}
            {M. Suchara et al.}

\normalsize\textlineskip
\thispagestyle{empty}
\setcounter{page}{1}

\vspace*{0.40truein}

\alphfootnote

\fpage{1}

\centerline{\bf
COMPARING THE OVERHEAD OF TOPOLOGICAL AND}
\vspace*{0.035truein}
\centerline{\bf CONCATENATED QUANTUM ERROR CORRECTION}
\vspace*{0.45truein}
\centerline{\footnotesize
MARTIN SUCHARA$^{1,}$\footnote{The corresponding author's email address is msuchar@us.ibm.com.}  , ARVIN FARUQUE$^2$, CHING-YI LAI$^3$,}
\centerline{\footnotesize
GERARDO PAZ$^3$, FREDERIC T. CHONG$^2$, JOHN KUBIATOWICZ$^4$}
\vspace*{10pt}
\vspace*{0.015truein}
\centerline{\footnotesize\it $^1$IBM Research, 1101 Kitchawan Road}
\baselineskip=10pt
\centerline{\footnotesize\it Yorktown Heights, NY 10598, U.S.A.}
\vspace*{10pt}
\vspace*{0.015truein}
\centerline{\footnotesize\it $^2$Computer Science Department, UC Santa Barbara, Harold Frank Hall}
\baselineskip=10pt
\centerline{\footnotesize\it Santa Barbara, CA 93106, U.S.A.}
\vspace*{10pt}
\vspace*{0.015truein}
\centerline{\footnotesize\it $^3$Department of Electrical Engineering, University of Southern California}
\baselineskip=10pt
\centerline{\footnotesize\it Los Angeles, CA 90089, U.S.A.}
\vspace*{10pt}
\vspace*{0.015truein}
\centerline{\footnotesize\it $^4$Computer Science Division, UC Berkeley, 673 Soda Hall}
\baselineskip=10pt
\centerline{\footnotesize\it Berkeley, CA 94720, U.S.A.}

\vspace*{0.35truein}

\abstracts{
This work compares the overhead of quantum error correction with concatenated and topological quantum error-correcting codes. To perform a numerical analysis, we use the Quantum Resource Estimator Toolbox (QuRE) that we recently developed. We use QuRE to estimate the number of qubits, quantum gates, and amount of time needed to factor a 1024-bit number on several candidate quantum technologies that differ in their clock speed and reliability. We make several interesting observations. First, topological quantum error correction requires fewer resources when physical gate error rates are high, white concatenated codes have smaller overhead for physical gate error rates below approximately $10^{-7}$. Consequently, we show that \emph{different} error-correcting codes should be chosen for two of the studied physical quantum technologies -- ion traps and superconducting qubits. Second, we observe that the composition of the elementary gate types occurring in a typical logical circuit, a fault-tolerant circuit protected by the surface code, and a fault-tolerant circuit protected by a concatenated code all differ. This also suggests that choosing the most appropriate error correction technique depends on the ability of the future technology to perform specific gates efficiently.
}{}{}

\vspace*{10pt}

\vspace*{9pt}

\vspace*{1pt}\textlineskip    
\input{introduction}
\input{ECCs}
\input{setup}
\input{results}

\input{conclusion}

\nonumsection{References}
\bibliography{RefEstimates}
\bibliographystyle{ieeetr}

\end{document}

%% file: introduction.tex
\section{Introduction}
\label{introduction}

In this paper we compare the overhead of quantum error correction in various realistic scenarios. In particular, we attempt to answer a question an experimentalist may have: ``Should my computer use a concatenated or topological error-correcting code?'' More generally, we provide the methodology and framework to systematically compare the performance of several competing designs in quantum computation from the viewpoint of  physical technology, algorithm, or error correction.

The performance of a quantum computer -- the number of qubits needed to run a specific quantum algorithm, its execution time, or the number of quantum gates that need to be applied -- is influenced by many factors. These include properties of the physical architecture, overhead of error-correcting codes, efficiency of decomposing arbitrary rotations into more elementary gates~\cite{depth_optimal_canonical}, the number of distilled magic states~\cite{BK:magicdist}, or the amount of communication between qubits. The contribution of these factors to the resource requirements and their interactions are sometimes difficult to predict, and an accurate performance estimate must take all of them into account. 

To facilitate numerical comparisons, we developed the Quantum Resource Estimator toolbox (QuRE)~\cite{QuRETechReport, QuREICCD}. The toolbox estimates resources such as number of qubits, running time or number of gates for a variety of candidate physical technologies, quantum algorithms, and quantum error-correcting codes. Unlike previous attempts to estimate resources which typically focused on one specific technology and one specific error-correcting code~\cite{layer, requirements_fault_factoring}, the QuRE toolbox is preloaded with examples of $12$ physical technologies, $7$ algorithms, and $4$ error-correcting codes, and is easily extendable to evaluate other choices.

To accurately estimate the overhead of quantum error correction, the QuRE toolbox heeds the locality constraints of quantum technologies -- two-qubit $CNOT$ gates can only be performed locally on two neighboring physical qubits. To that end, we use a tiled qubit layout for concatenated codes. Each tile contains physical qubits that represent the state of a single fault-tolerant logical qubit. To perform $CNOT$ gates inside each tile, either $SWAP$ gates or ballistic movement must be used to move the two interacting qubits together. Optimized tiled qubit layouts and movements for the Steane code, Bacon-Shor code, and the Knill's postselection scheme are designed in~\cite{noise_threshold_fault, latency_local_2d, knill_postselection_architecture}. The QuRE toolbox simulates this movement. QuRE also uses a tiled qubit layout for the surface code, where a pair of holes, representing a logical qubit, resides inside a tile. However, the computation and error correction with the surface code are inherently local, and no swapping of qubits is needed.

To simplify the presentation, we selected the Bacon-Shor and surface codes as representatives for concatenated and topological codes, and we analyze results for three quantum technologies that vary in gate times and reliability. The tradeoff of clock speed and reliability is seen on the supecronducting technology~\cite{quantum_information_processing} which has a fast clock cycle but relatively unreliable gates, and the slower but more reliable ion trap technology~\cite{information_processing_trapped}. To provide numerical examples, we chose Shor's algorithm~\cite{polynomial_factorization}. This is the best known quantum algorithm and a 1024-bit length of the product is a popular choice in the literature that allows comparison of our results~\cite{layer}.

In this paper we make the following observations:
\begin{itemize}
  \item With increasing error rate of the physical technology, the resource requirements of topological codes increase much more gradually than those of concatenated codes. 
  \item For a typical real-world sized quantum algorithm with $1 \times 10^{15}$ logical gates, the surface code has lower overhead than the Bacon-Shor code when the physical error rate is above approximately $1 \times 10^{-7}$, and vice versa.
  \item The composition of the quantum gates that need to be applied differs based on the choice of the error-correcting code. The gates that appear in a typical non-fault-tolerant logical circuit are also different. Whereas logical circuits frequently use $S$ and $T$ gates, the dominant gates in fault-tolerant circuits are $SWAP$ for concatenated codes and $CNOT$ for the surface code.
  \item The optimal choice of error-correcting codes depends on the error properties of the technology and its ability to perform $CNOT$ and $SWAP$ gates efficiently.
\end{itemize}

The paper is organized as follows. Section~\ref{ECCs} introduces the Bacon-Shor and surface quantum error-correcting codes, and describes how the QuRE toolbox estimates their overhead. Section~\ref{setup} describes the properties of three specific physical technologies, properties of the Shor's factoring algorithm, and explains how the QuRE toolbox uses these inputs to perform resource estimation. In Section~\ref{results} we show our numerical results, which include the resources needed for factoring a 1024-bit number, and and then we compare the performance of the Bacon-Shor and surface code. In Section~\ref{results} we also discuss the limitations of our methodology.

%% file: ECCs.tex
\section{Quantum Error Correction Overhead}
\label{ECCs}

To compare the overhead of concatenated and topological quantum error-correcting codes, we choose the Bacon-Shor~\cite{AC:baconshor} and surface~\cite{topological_quantum_memory} codes as representatives. Here we briefly describe the properties of the two codes and how the QuRE toolbox estimates their performance.

\subsection{The Bacon-Shor Code}
\label{BaconShor}

The Bacon-Shor code uses nine unreliable qubits to encode a single more reliable qubit. The code can be concatenated. Each encoded qubit block consists of nine lower-level blocks. \emph{Logical} qubits are defined as the fault-tolerant qubits built of a greater number of unreliable physical qubits, and \emph{logical gates} are reliable quantum operations applied to the logical qubits. The Bacon-Shor code belongs to the family of Calderbank-Shor-Steane (CSS) codes. Most logical gates can be performed transversally, meaning that the operation is applied nine times to each of the lower level blocks. The $S$ and $T$ gates are non-transversal. High-quality ancillas must be distilled to perform the $S$ and $T$ operations, and this requires additional resources.

\begin{figure}[t!]
\centering
\includegraphics[width=.75\textwidth]{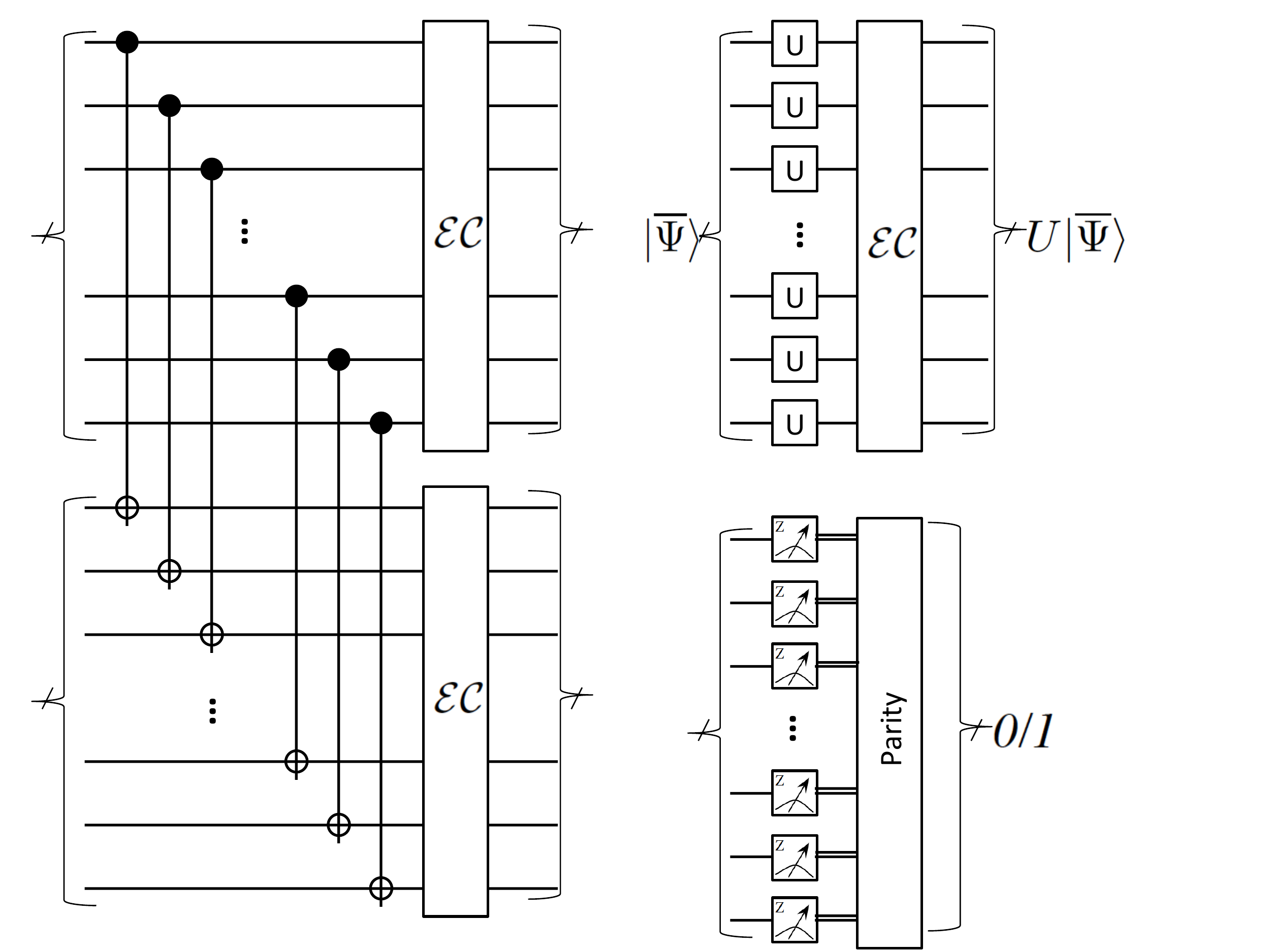}
\fcaption{Fault-tolerant implementation of the $CNOT$ gate (on the left), the single qubit gates $X$, $Y$, $Z$ and $H$ (top right), and $Z$ basis measurement (bottom right).}
\label{fig:transversalGates}
\end{figure}

The transversal implementations of the Pauli gates, the H (Hadamard) gate, measurements, and the $CNOT$ (controlled not) gate are shown in Fig.~\ref{fig:transversalGates}. Estimating the cost of transversal single-qubit gates is straightforward. The cost of a two-qubit $CNOT$ gate depends on the layout of the physical qubits in the architecture because a physical $CNOT$ gate can be only applied to adjacent qubits. 

\begin{figure}[t!]
\centering
\includegraphics[width=0.25\textwidth]{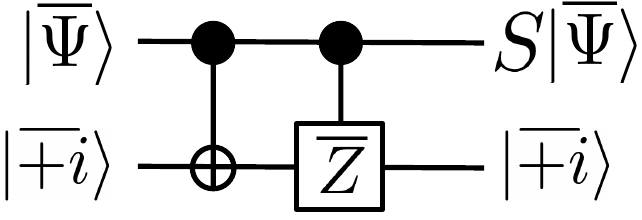}
\fcaption{Application of the S gate requires an ancilla in the $\ket{+i}$ state.}
\label{fig:sgate}
\end{figure}

\begin{figure}[t!]
\centering
\includegraphics[width=0.9\textwidth]{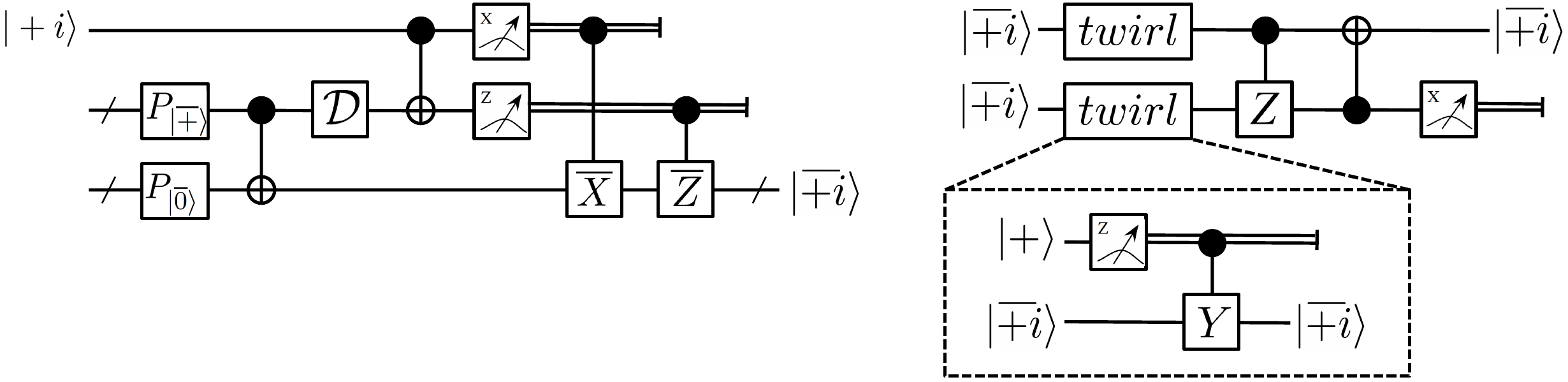}
\fcaption{To distill an ancilla in the $\ket{+i}$ state, the we first use the circuit on the left to inject the state into the code space. Then the circuit on the right is used repeatedly to distill a state with sufficient fidelity.}
\label{fig:idist}
\end{figure}

The $S$ gate can be applied using the circuit in Fig.~\ref{fig:sgate}. It uses an ancilla in the state $\ket{{+i}} = \frac{\ket{0} + i \ket{1}}{\sqrt{2}}$ to generate the required gate. In turn, an encoded $\ket{\overline{+i}}$ state at any level of concatenation can be obtained via the injection circuit on the left side in Fig.~\ref{fig:idist}, which teleports an arbitrary lower-level state, in this case $\ket{+i}$, into an encoded state $\ket{\overline{+i}}$ at the cost of decoding (depicted as $\mathcal{D}$ in the circuit) an encoded Bell pair. Moreover, since the injection circuit is not fault-tolerant, a higher fidelity $\ket{\overline{+i}}$ has to be distilled via multiple {\it successful} rounds of the circuit on the right side of Fig.~\ref{fig:idist}.

In order to have a universal gate set, we also need the $\pi/8$ gate, called the $T$ gate. A fault-tolerant version of this gate also cannot be constructed transversally. Construction of the $T$ gate requires distillation of a special ancillary state $T\ket{+}$ with sufficient fidelity. The process is similar to the distillation process of the $\ket{\overline{+i}}$ state, and we describe it in~\cite{QuRETechReport}.

The error correction operation $\mathcal{EC}$ must be performed after applying each gate, as well as periodically on idle qubits. Steane's syndrome extraction~\cite{efficient_fault_tolerant, quantum_accuracy_threshold} is used. Each syndrome measurement requires preparation of ancillas in a specific state, application of a number of $CNOT$ gates, qubit measurements, and a classical parity calculation. To correct errors, the X and Z gates are applied according to the syndrome.

Since the two-qubit $CNOT$ gates can be only applied to \emph{physically adjacent} qubits, we must take the cost of communication into account. We assume that the state of one of the two interacting qubits is transferred using a sequence of $SWAP$ operations before applying the $CNOT$. Moreover, additional ``dummy'' qubits must be introduced because two data qubits cannot be swapped directly. Spedalieri et al. describe optimized tiled qubit operations for the Bacon-Shor code to minimized the overhead of communication ~\cite{latency_local_2d}. We use their optimized operations. Each logical qubit is represented by a two-dimensional tile of $7$ by $7$ smaller tiles at the next lower concatenation level. It follows that an algorithm with $n$ logical qubits requires $n(7 \times 7)^l$ physical qubits with $l$ levels of concatenation.

The concatenation level $l$ of the code must be sufficient to correct errors with high probability. QuRE determines $l$ such that the probability of failure of the quantum computation due to uncorrected errors is less than approximately $50\%$. An estimation method from~\cite{book_nielsen_chuang} is used. Let $p$ be the failure probability of a physical gate. The probability that a circuit introduces two errors, which is an unrecoverable error, is $O(p^2) = cp^2$. Here $c = 1/p_{th} = 1 / 2.02 \times 10^{-5}$ is the reciprocal of the error threshold~\cite{noise_threshold_fault}. With $l$ levels of concatenation the failure probability becomes $(cp)^{2^{l}}/c$. With $N$ gates  and error probability at most $0.5$, each gate must be accurate to $0.5/N$ so it suffices to find $l$ satisfying: $\frac{(cp)^{2^{l}}}{c} \leq \frac{\epsilon}{N}$.

It is convenient to use recursion to express the cost of a gate at the $m$-th concatenation level. The QuRE toolbox follows the optimized movement strategy described by Spedalieri et al. \cite{latency_local_2d} and counts the number of gates required by each operation. The simplest example of this recursion expresses the number of gates required by the $X$ operation: $ops(X_{(m)}) = 9 ops(X_{(m-1)}) + ops(\mathcal{EC}_{(m)})
$, where $\mathcal{EC}_{(m)}$ represents all gates required by the error correction circuit. The gate time is $time(X_{(m)}) = time(X_{(m-1)}) + time(\mathcal{EC}_{(m)})$ because the nine $X$ operations can be performed in parallel. The cost of other gates is estimated similarly.

Our resource estimation uses an ancilla factory model. Sections of the quantum computer are devoted to producing and distilling the special ancillas required by the $S$ and $T$ gates. QuRE estimates a sufficient number of distillation rounds so that the error of the distilled state is lower than the error of reliable Clifford gates protected by the Bacon-Shor code. We found that $3$ to $5$ distillation rounds are sufficient. The cost of distillation is added to the resource estimate.

\subsection{The Surface Code}
\label{ECCs:Surface}

\begin{figure}
\centering
\begin{minipage}{.46\textwidth}
  \centering
  \includegraphics[width=1\linewidth]{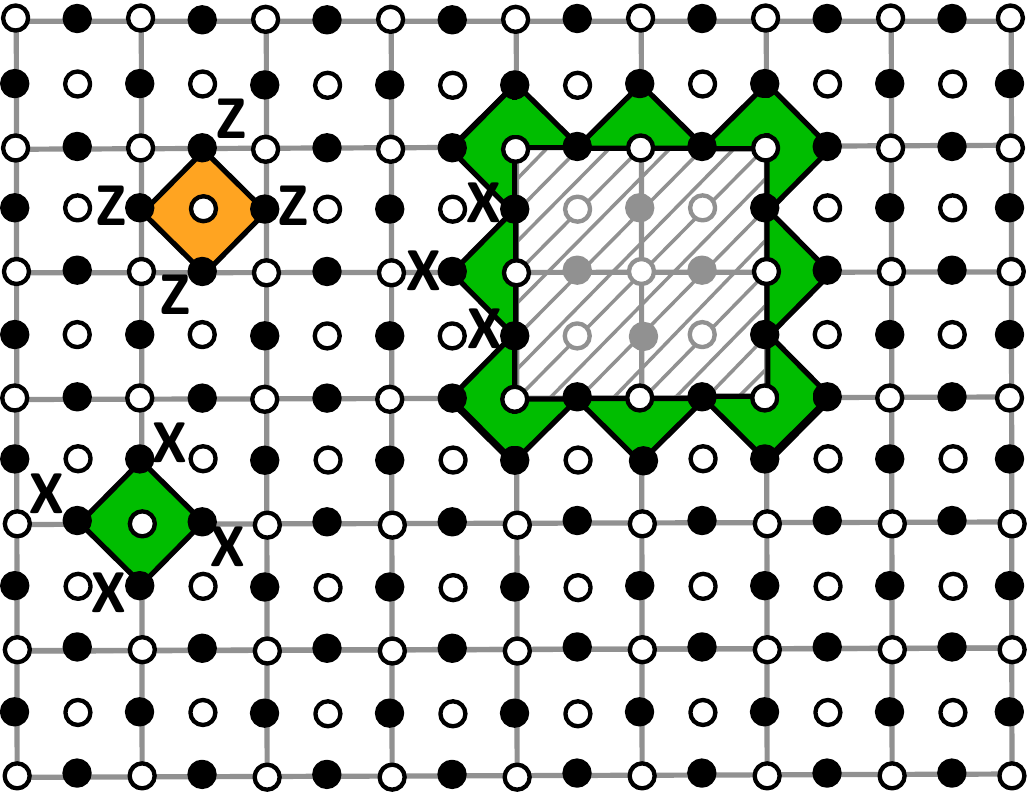}
  \fcaption{Lattice of the surface code. A few syndromes and a smooth hole are shown.}
  \label{fig:surfaceLattice}
\end{minipage}
\hfill
\begin{minipage}{.46\textwidth}
  \centering
  \includegraphics[width=1\linewidth]{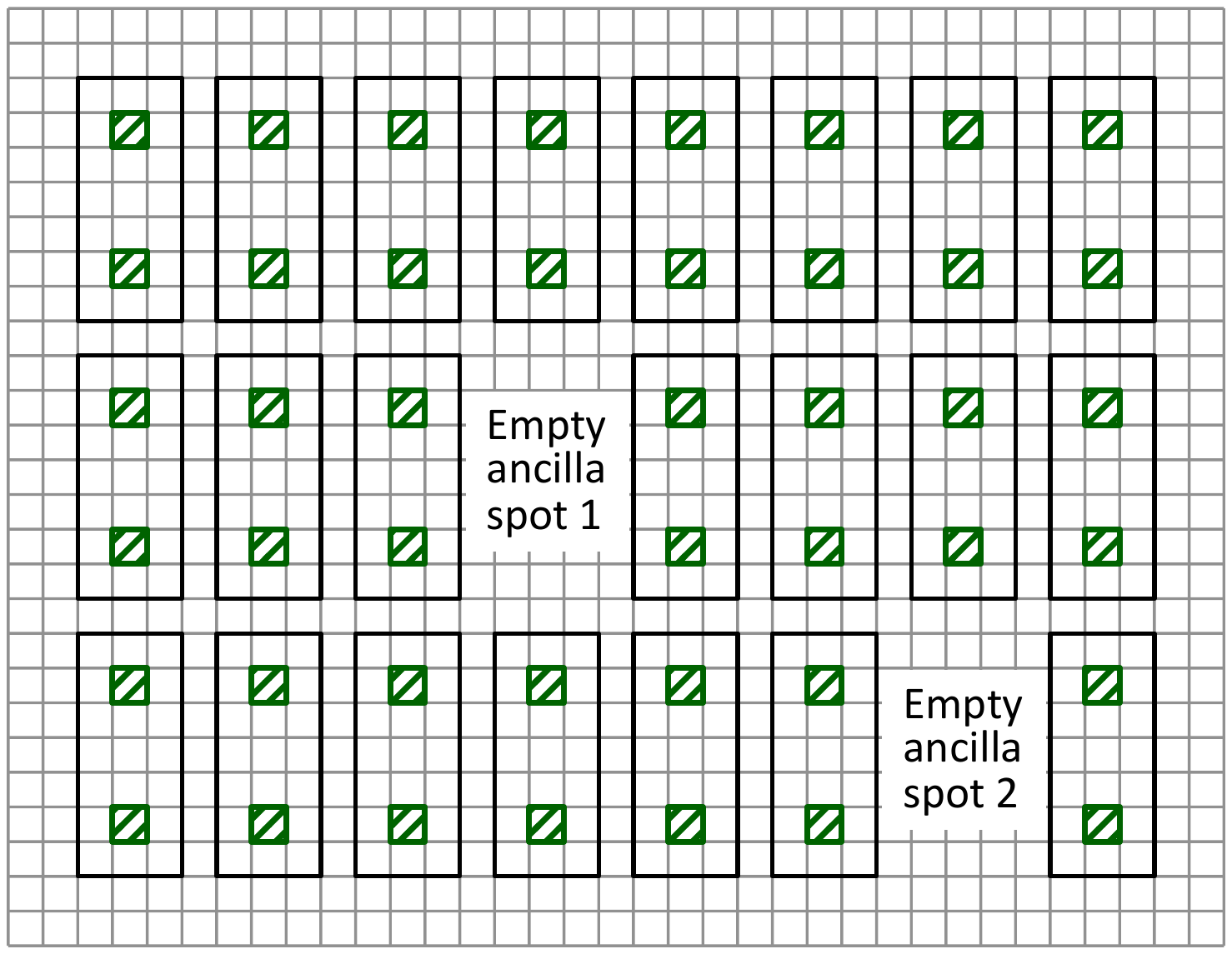}
  \fcaption{The tiled layout of the surface code. Tiles contain two smooth holes (dashed).}
  \label{fig:surfaceTiles}
\end{minipage}
\end{figure}

The surface code~\cite{topological_quantum_memory} places qubits on a regular grid, such as the one shown in Fig.~\ref{fig:surfaceLattice}. An advantage of the code is locality -- all operations can be performed locally without the need to move qubits. Another advantage of the code is that it can tolerate higher error rates than concatenated codes.

\begin{figure}[t!]
\centering
\includegraphics[width=.60\textwidth]{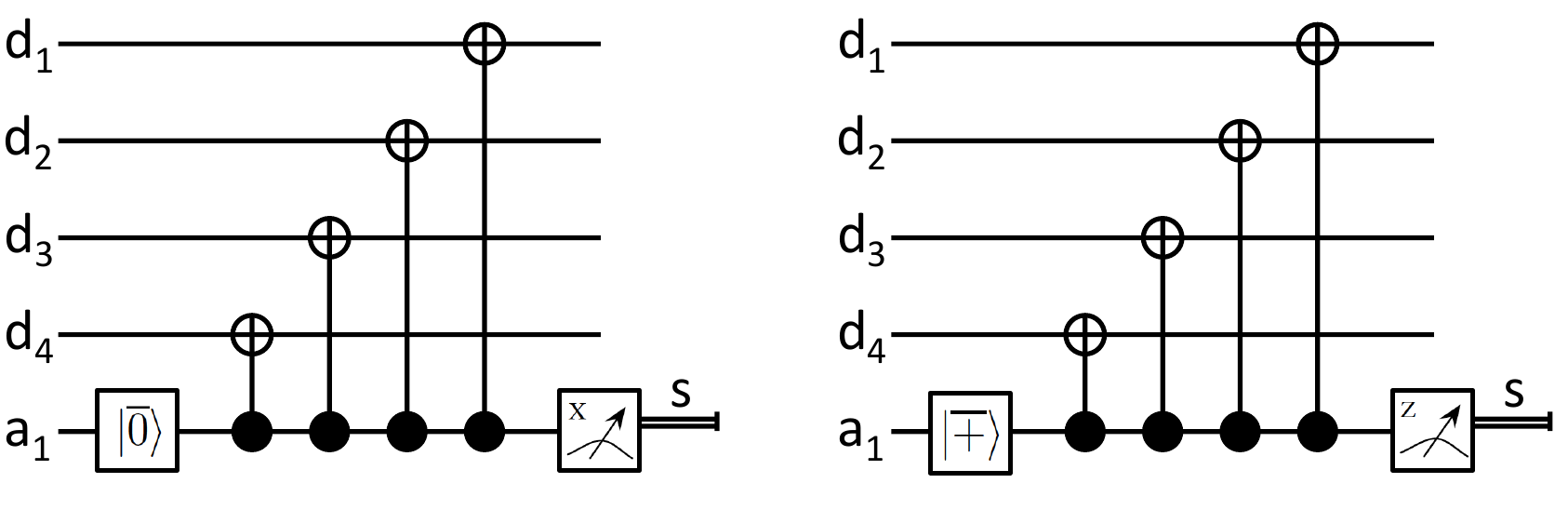}
\fcaption{The syndrome measurement circuits. Initialization, four CNOTs and a measurement are used to obtain syndrome.}
\label{fig:SteaneMeas}
\end{figure}

Fig.~\ref{fig:surfaceLattice} shows the qubit layout of the surface code simulated by QuRE. The black circles represent data qubits that encode the quantum state, and the white qubits represent ancillas that are used for error syndrome extraction needed to diagnose errors. The syndromes are determined by measuring type $XXXX$ and $ZZZZ$ stabilizers, a few of which are depicted in Fig.~\ref{fig:surfaceLattice}.
Syndrome measurement is the most expensive operation because it is performed continuously in the entire surface. The $XXXX$ and $ZZZZ$ syndromes are extracted using the circuits in Fig.~\ref{fig:SteaneMeas}, each of which requires one ancilla. There are enough ancillas in Fig.~\ref{fig:surfaceLattice} to allow simultaneous measurement of all syndromes.

Holes in the surface are regions where the stabilizers are not measured. When no holes are present, the surface code only encodes two logical qubits~\cite{topological_quantum_memory}. To increase the number of encoded qubits, we need to introduce new holes into the surface.  One additional logical qubit is encoded by introducing a \emph{pair} of smooth or rough holes.  Fig.~\ref{fig:surfaceLattice} shows an example of a single ``smooth'' hole (dashed). The hole is surrounded by $XXX$ stabilizers. ``Rough'' holes can be introduced as well, they are shifted by half a cell size, and are surrounded by $ZZZ$ stabilizers. To count the number of physical qubits, we introduce a tiled qubit layout. Each tile contains a pair of rough holes, and therefore represents one logical qubit. Additional tiles are needed for ancillas as shown in Fig.~\ref{fig:surfaceTiles}.

\begin{figure}[t!]
\centering
\includegraphics[width=.60\textwidth]{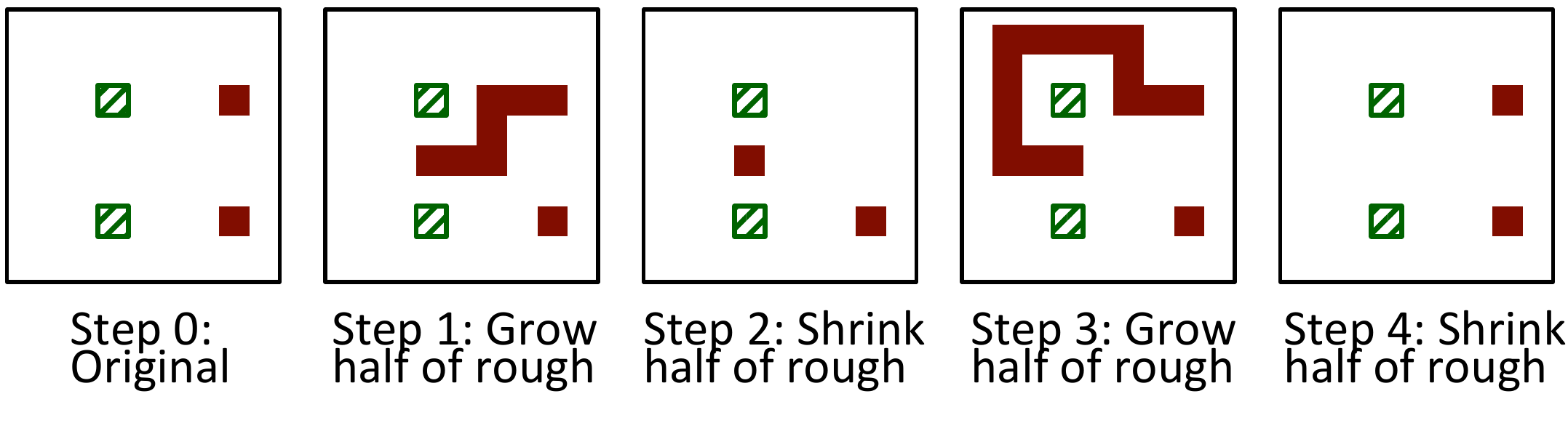}
\fcaption{A smooth-rough $CNOT$ gate. The time evolution of the smooth holes (dashed fill) and rough holes (solid fill) is shown.}
\label{fig:smoothRoughCnot}
\end{figure}

\begin{figure}[t!]
\centering
\includegraphics[width=.55\textwidth]{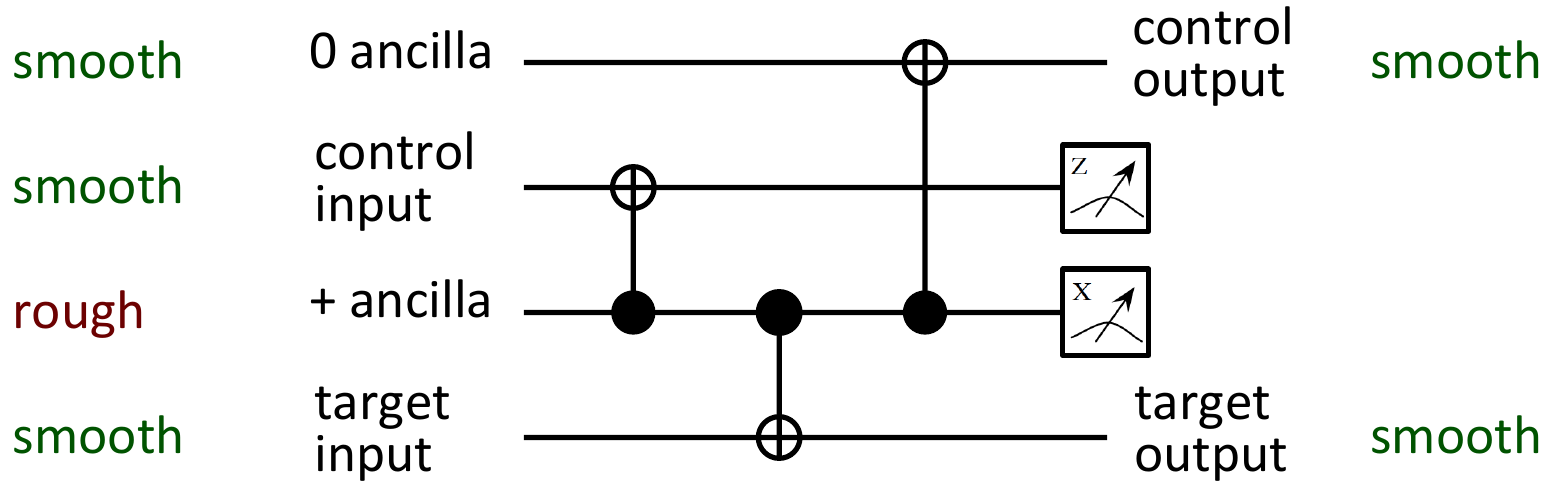}
\fcaption{A smooth-smooth $CNOT$ gate uses three smooth-rough CNOTs.}
\label{fig:smoothSmoothCnot}
\end{figure}

Interestingly, the surface code does not need any $SWAP$ operations to perform a $CNOT$ gate between logical qubits, even if they are far apart. $CNOT$ operations can be done easily between smooth and rough hole pairs by a braiding procedure where one of the rough holes is ``grown'' and ``shrunk'' to move around a smooth hole in the other hole pair~\cite{surfaceCode}. This operation is illustrated in Fig.~\ref{fig:smoothRoughCnot}. Since our tiled layout only contains smooth holes, we need to estimate the cost of a smooth-smooth $CNOT$ between two smooth hole pairs. This operation is a bit more complicated and requires additional ancilla space to perform smooth to rough qubit conversion. Fig.~\ref{fig:smoothSmoothCnot} illustrates this. First the two ancillas are initialized, and then three $CNOT$s are performed by braiding.

Similarly to the Bacon-Shor code, we require that the failure probability of the computation is below $50\%$. To achieve this, we need to choose a sufficiently large code distance $d$ (the number of qubits between the closest pair of holes in Fig.~\ref{fig:surfaceTiles}). We use the method of Jones~\cite{latency_local_2d}. An estimate of $d$ is obtained by solving $\frac{0.5}{N} \ge C_1 \left (C_2 \frac{p}{p_{th}} \right )^{\lfloor \frac{d + 1}{2} \rfloor}$ for $d$, where $0.5/N$ is the desired success probability divided by the number of logical gates in the algorithm, $p$ is the physical gate error rate, $p_{th} \approx 0.01$ is the threshold of the surface code, and $C_1 \approx 0.13$ and $C_2 \approx 0.61$ are constants.

The QuRE toolbox estimates resources in three key steps. First the number of physical qubits is calculated. Second, the actual time needed to run the entire algorithm is determined. Third, the number of physical gates is estimated.
Physical qubits are calculated as follows. First, we obtain the number of tiles by adding the number of logical qubits, ancillas to support $CNOT$ operations, and additional space for magic state distillation. We multiply this number by the number of physical qubits needed to build a single tile.
The total running time is obtained by adding the cost of all elementary logical operations. Here we derive the running time of the logical $CNOT$. For other operations see~\cite{QuRETechReport}.

To estimate the running time of a logical $CNOT$ gate, we first need to consider more elementary operations. To perform error correction ($\mathcal{EC}$), all syndromes must be measured $d$ times. The syndromes can be measured in parallel using the circuit in Fig.~\ref{fig:SteaneMeas}, and therefore the time to perform the error correction $time(\mathcal{EC})$ is equal to $d$ times the sum of the time to perform the state preparations, $4$ $CNOT$ gates, and the $X$ and $Z$ measurements. The $CNOT$ gate between a smooth and rough qubit can be performed in time $4 \times time(\mathcal{EC})$ by the braiding procedure shown in Fig.~\ref{fig:smoothRoughCnot} because the error correction operation must be applied after each hole expansion or contraction. A smooth logical qubit can be measured in the $Z$ basis and a rough qubit can be measured in the $X$ basis, and each operation takes the time of a physical measurement plus one error correction $\mathcal{EC}$.
Finally, the time needed to perform a smooth-smooth $CNOT$ gate, shown in Fig.~\ref{fig:smoothSmoothCnot}, must be the sum of the time of three smooth-rough $CNOT$s and the two measurements done in parallel (the ancilla state preparations are performed offline): $time(CNOT) = 3 \times 4 \times time(\mathcal{EC}) + time(\mathcal{M}) + time(\mathcal{EC})$ where $time(\mathcal{M})$ represents a state measurement.

QuRE obtains the total physical gate count in two steps. First, an estimate is obtained by calculating the number of gates needed to continuously perform the error correction operation in the entire surface of the code for the time period needed to run the algorithm. This is done by multiplying three terms: the number of error correction cycles, the number of elementary cells in the surface, and the number of operations needed to do an error correction for an elementary cell. The last term simply consists of the gates in the two syndrome extraction circuits of Fig.~\ref{fig:SteaneMeas}. In the second step, a refinement is performed by including the small number of additional gates needed by logical operations.

%% file: setup.tex
\section{Simulation Setup}
\label{setup}

Here we describe how we used the QuRE toolbox to obtain the numerical results in Section~\ref{results}. A more detailed description of the toolbox is in~\cite{QuRETechReport, QuREICCD}.

\subsection{Numerical Simulations with QuRE}
\label{setup:toolbox}

\begin{figure*}[b!]
\centering
\includegraphics[width=0.75\textwidth]{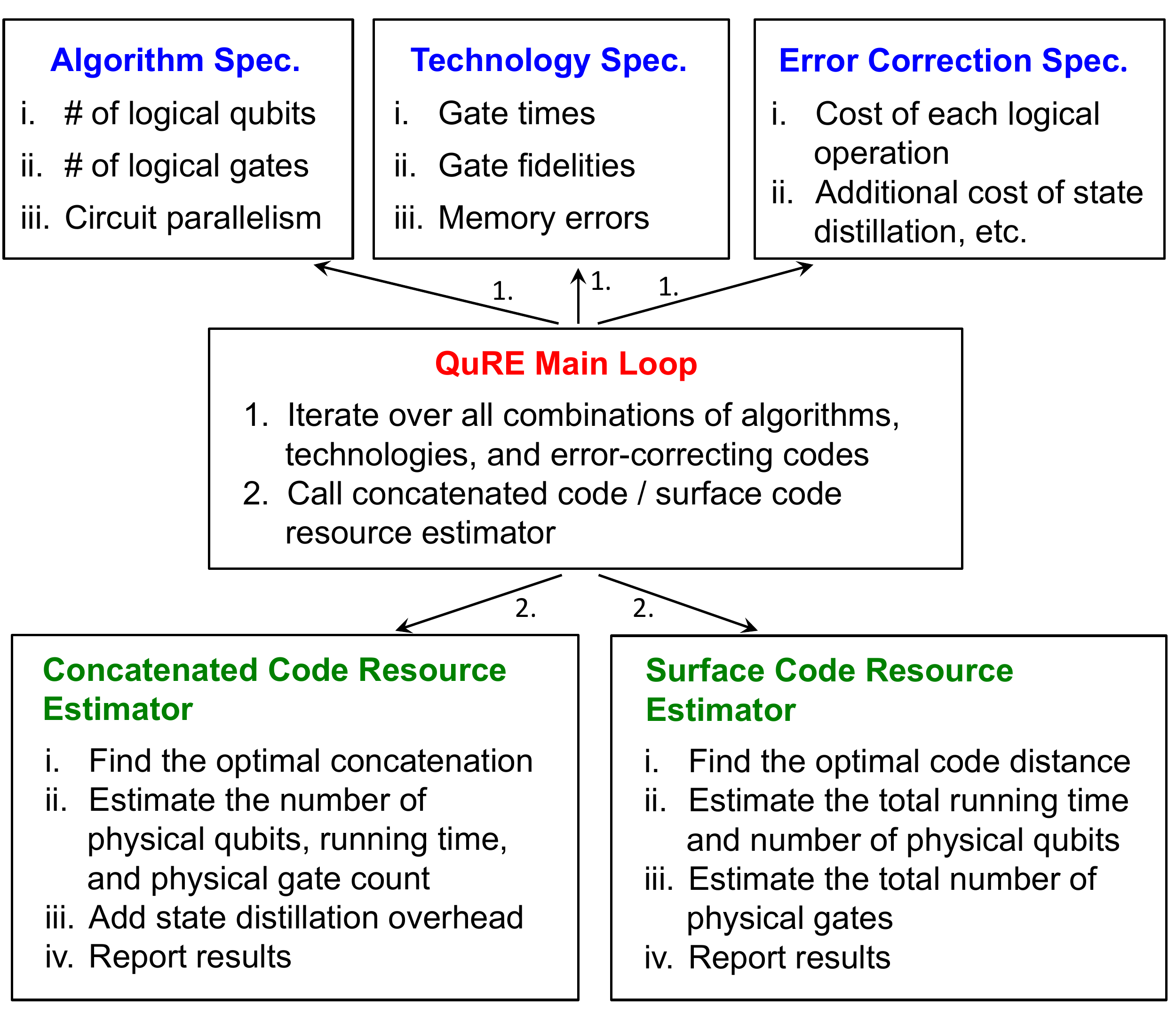}
\fcaption{Key Modules of the QuRE Toolbox.}
\label{fig:toolFlow}
\end{figure*}

Fig.~\ref{fig:toolFlow} shows a schematic view of the QuRE toolbox. At the heart of the tool is the \emph{Main Loop}, which iterates over all specified quantum algorithms, quantum technologies, and quantum error-correcting codes and calls all other modules.

A \emph{Technology Specification} module describes properties of a particular physical quantum technology. It specifies the time needed to carry out each physical gate, the error rate of the worst gate, and information about memory error rate per unit time. Numeric characterization of the three quantum technologies used in this work are in Subsection~\ref{setup:technologies}.

An \emph{Algorithm Specification} module provides information about the number of logical qubits a particular algorithm needs. Number of logical gates and simplified information about the circuit parallelism are also specified by the module. Numeric inputs that we used are summarized in Subsection~\ref{setup:algorithm}.

An \emph{Error Correction Specification} module is provided for each supported quantum error-correcting code. It quantifies the time and the number of physical gates needed to implement a logical gate of each type at an arbitrary level of concatenation (or, in case of topological codes, for an arbitrary code distance), as described in Section~\ref{ECCs}.

The \emph{Concatenated Code Resource Estimator} and the \emph{Surface Code Resource Estimator} modules report the resources needed by the specific algorithm, technology, and concatenated code loaded by the Main Loop. The module determines the minimum concatenation level or code distance that is sufficient to complete the algorithm successfully with high probability, and then estimates the cost of performing all logical operations fault-tolerantly. 

\subsection{Physical Quantum Technologies}
\label{setup:technologies}

We use models of quantum technologies and quantum control protocols that were studied by Hocker et al.~\cite{analysis_PMDs}. QuRE includes all twelve technologies from~\cite{analysis_PMDs}, and here we introduce three of them. These three technologies are among the most promising candidates suitable for building a large-scale quantum computer, representing a range of properties that a future quantum computer may possess -- very fast but error prone superconductors, slower but more reliable ion traps, and neutral atoms with only average speed and average error properties.

\textbf{Superconductors~\cite{quantum_information_processing, superconducting_phase_qubits}:}
The building block for qubits is the Josephson junction. This technology suffers from relatively high errors due to radiation leakage into the Josephson junction, circuit defects, and engineering limitations~\cite{analysis_PMDs}.

\textbf{Ion Traps~\cite{information_processing_trapped}:}
This technology is based on a 2D lattice of ions confined by electromagnetic field. Lasers are applied to implement quantum gates. Errors are caused by intensity fluctuations of the laser, resulting in low gate errors.

\textbf{Neutral Atoms~\cite{quantum_information_rydberg}:}
Qubits are represented by ultracold atoms trapped by light waves in an optical lattice. The ``ultracold'' atom properties improve their noise resilience, but errors arise from atomic motion inside the lattice.

The durations of the physical gates are shown in Table~\ref{table:PMDGates}.  The leftmost column shows the one- and two-qubit gates, measurements and state preparations that can be performed. Note that some gates can be constructed from more elementary operations in the quantum hardware. For example, the $SWAP$ gate could be constructed using three $CNOT$ gates. Reliability is another important property of the technologies. Table~\ref{table:PMDErrors} summarizes the error probability after applying the worst gate, as well as the probability of a bit flip per nanosecond on an idle qubit.

The models of Hocker et al.~\cite{analysis_PMDs} consider many details that a realistic computer must posses, including a basic instruction set, errors due to qubit movement and decoherence, and the use of currently known control protocols to optimize properties of quantum gates. While the experimental demonstrations of these technologies to date have been limited to a small scale and achieved error rates are typically worse, the parameters of the models used here may be achievable in the future.

\begin{table}[t!]
\tcaption{The gate times (in ns) for all supported one- and two-qubit quantum gates.}
\ra{1.1}
\begin{center}
\small
\begin{tabular}{llll}\toprule
\textbf{Gate}                          &\textbf{Superconductors}        &\textbf{Ion Traps}        & \textbf{Neutral Atoms}\\ 
\toprule
CNOT                       & $22$             & $120,000$                & $11,370$\\
SWAP                       & $17$             & $10,000$                  & $34,120$\\
H                               & $6$                & $6,000$                    & $2,991$\\
$\Ket{+}$ prep.        & $100$           & $16,000$                  & $3,991$\\
$\Ket{0}$ prep.        & $106$           & $10,000$                  & $1,000$\\
X meas.                    & $16$             & $106,000$                & $82,991$\\
Z meas.                    & $10$             & $100,000$                & $80,000$\\
X                               & $10$             & $5,000$                     & $2,667$\\
Y                               & $10$             & $5,000$                     & $2,667$\\
Z                               & $1$               & $3,000$                     & $5,532$\\
S                               & $1$               & $2,000$                     & $3,125$\\
T                               & $1$               & $1,000$                     & $3,125$\\
\bottomrule
\end{tabular}
\label{table:PMDGates}
\end{center}
\end{table}

\begin{table}[t!]
\tcaption{The probability of error of the worst gate and the probability of an error occurring on an idle qubit per nanosecond for the three quantum architectures.}
\ra{1.1}
\begin{center}
\small
\begin{tabular}{llll}\toprule
\textbf{Error}                      &\textbf{Superconductors}        &\textbf{Ion Traps}                   & \textbf{Neutral Atoms}\\ 
\toprule
Gate                                    & $ 1.00 \times 10^{-5}$             & $ 3.19 \times 10^{-9}$         & $ 1.47 \times 10^{-3}$\\
Memory                              &  $1.00 \times 10^{-5}$              & $2.52 \times 10^{-12}$       & not available\\
\bottomrule
\end{tabular}
\label{table:PMDErrors}
\end{center}
\end{table}

\subsection{Shor's Factoring Algorithm}
\label{setup:algorithm}

In order to compare the properties of quantum error-correcting codes, we need to use a representative quantum algorithm with the right ``mix'' of quantum gates and known parallelism properties. For this purpose, we choose the Shor's factoring algorithm~\cite{polynomial_factorization} for factoring a $1024$-bit number. One variant of the algorithm requires approximately $1.68 \times 10^8$ Toffoli gates and $6,144$ logical qubits~\cite{layer}. The number of gates of other types is negligible. We obtained an approximate logical gate count by decomposing the Toffoli gates into more elementary $CNOT$, $H$, $T$, and $T^{\dagger}$ logical gates using the decomposition from~\cite{algorithm_T_count}. The approximate gate count is shown in Table~\ref{table:AlgorithmGates}. Note that these tables summarize the \emph{logical} resource requirements of the algorithms before the overhead of error correction is taken into account. The parallelism estimate shown in Table~\ref{table:AlgorithmGates} is a conservative estimate of how many of the gates can be performed in parallel.

\begin{table}[t!]
\tcaption{Logical gate count for Shor's algorithm factoring a 1024-bit number. A conservative estimate of parallelization factors shown.} 
\ra{1.1}
\begin{center}
\small
\begin{tabular}{lllllll}\toprule
\textbf{Gate}                                       & \textbf{Occurences} & \textbf{Parallelism}\\
\toprule
$CNOT$                                             & $1.18 \times 10^{9}$         & $1$\\
$H$                                                     & $3.36 \times 10^{8}$          & $1$\\
$T$ or $T^{\dagger}$                           & $1.18 \times 10^{9}$         & $2.33$\\
\bottomrule
\end{tabular}
\label{table:AlgorithmGates}
\end{center}
\end{table}

%% file: results.tex
\section{Numerical Comparisons of Concatenated and Topological Codes}
\label{results}

Subsection~\ref{results:Shor} compares the resources required to run Shor's factoring algorithm with the Bacon-Shor and surface codes. Subsection~\ref{results:technologies} considers the error correction overhead of the two codes as a function of the physical gate error rate of an abstract quantum technology. Subsection~\ref{results:gates} compares the gate composition of a typical logical circuit and fault-tolerant circuits. Finally, Subsection~\ref{results:limitations} discusses the limitations of our methodology.

\subsection{Comparisons for Shor's Algorithm}
\label{results:Shor}

\begin{table}[b!]
\tcaption{The resources needed to factor a 1024-bit number with Shor's algorithm. Results shown for the surface and Bacon-Shor codes on three technologies.} 
\ra{1.1}
\begin{center}
\small
\begin{tabular}{lllll}\toprule
\multirow{2}{*}{\textbf{Technology}} & {\textbf{Neutral}} & \textbf{Supercond.} & \textbf{Ion} & \\
& \textbf{Atoms} & \textbf{Qubits} & \textbf{Traps} & \\
\textbf{Gate error}                         & $\bm{1 \times 10^{-3}}$ & $\bm{1 \times 10^{-5}}$ & $\bm{1 \times 10^{-9}}$ & \\
\textbf{Avg. gate time}                  & \textbf{19,000 ns} & \textbf{25 ns} & \textbf{32,000 ns} & \\
\toprule
\textbf{Execution time}                 & $2.62$ years                        & $10.81$ hours                      & $2.22$ years & \\
\textbf{No. qubits}                         & $5.29 \times 10^{8}$            & $4.57 \times 10^{7}$           & $1.44 \times 10^{8}$ & \\
\textbf{No. gates}                          & $1.02 \times 10^{21}$          & $2.55 \times 10^{19}$        & $5.10 \times 10^{19}$ & \\
\textbf{Dominant gate}                 & $CNOT$                                 & $CNOT$                               & $CNOT$ & \\
\textbf{Code distance}                & $17$                                        & $5$                                        & $3$                                 &        \textbf{Surface code} \\
\textbf{Logical gate error}            & $4.99 \times 10^{-11}$        & $2.95 \times 10^{-11}$       & $4.92 \times 10^{-15}$ & \\
\textbf{Logical gate time}             & $1.29 \times 10^{5}$ ns       & $2.10 \times 10^{2}$ ns    & $5.96 \times 10^{5}$ ns & \\
\textbf{No. qubits per logical}      & $3.73 \times 10^{4}$           & $3.23 \times 10^{3}$          & $1.16 \times 10^{3}$ & \\
\textbf{No. gates per logical}       & $1.11 \times 10^{5}$           & $9.60 \times 10^{3}$          & $3.46 \times 10^{3}$ & \\
\toprule
\textbf{Execution time}                 & N/A                                          & $5.10$ years                          & $57.98$ days & \\
\textbf{No. qubits}                         & N/A                                          & $2.65 \times 10^{12}$           & $4.60 \times 10^{5}$ & \\
\textbf{No. gates}                          & N/A                                          & $1.16 \times 10^{32}$           & $4.07 \times 10^{18}$ & \\
\textbf{Dominant gate}                 & N/A                                          & $SWAP$                                  & $CNOT$ & \\
\textbf{Code concatenations}      & N/A                                        & $5$                                           & $1$                                  &        \textbf{Bacon-Shor code} \\
\textbf{Logical gate error}            & N/A                                          & $3.42 \times 10^{-15}$         & $5.09 \times 10^{-14}$ & \\
\textbf{Logical gate time}             & N/A                                          & $1.42 \times 10^{7}$ ns        & $7.27 \times 10^{5}$ ns & \\
\textbf{No. qubits per logical}      & N/A                                          & $2.82 \times 10^{8}$            & $49$ & \\
\textbf{No. gates per logical}       & N/A                                          & $1.18 \times 10^{11}$          & $79$ & \\
\bottomrule
\end{tabular}
\label{table:ECCComparison}
\end{center}
\end{table}

Table~\ref{table:ECCComparison} shows the resources needed by Shor's algorithm to factor a 1024-bit number. The following quantities are shown:
\begin{itemize}
\item \textbf{Execution time:} the total time needed to obtain the correct result with high probability.
\item \textbf{No. qubits:} the total number of physical qubits needed to build the quantum computer.
\item \textbf{No. gates:} the total number of gates executed by the quantum computer.
\item \textbf{Dominant gate:} the most frequently occurring gate.
\item \textbf{Code distance or code concatenations:} the distance for the surface code or the number of concatenations for the Bacon-Shor code.
\item \textbf{Logical gate error:} the error probability after performing one logical operation on a single logical qubit.
\item \textbf{Logical gate time:} the time in ns to perform the error-correcting operation on one logical qubit.
\item \textbf{No. qubits per logical:} the size of a tile that stores one logical qubit.
\item \textbf{No. gates per logical:} the number of gates required to perform the error-correcting operation on the tile.
\end{itemize}

The data in Table~\ref{table:ECCComparison} demonstrates some key differences between the Bacon-Shor and surface codes. For the superconducting technology, the surface code offers running time on the order of hours whereas the Bacon-Shor code requires several years to completion. In case of ion traps, the situation reverses, and the concatenated Bacon-Shor code is much more efficient than the surface code. The reason for this is that the Bacon-Shor code is very efficient with just $1$ or $2$ concatenations, and therefore works well on ion traps which have a very low gate error rate. On the other hand, the surface code works comparably well for any code distance (i.e., for any gate error rate sufficiently below the threshold of the code), but the long gate time of ion traps make the use of the code impractical. These observations are further supported by results that we present next. Note that for technologies with high error rates, such as the neutral atom technology, only the surface code can be used, as the error correction threshold of the Bacon-Shor code is not met.

\subsection{Comparisons for Different Technologies}
\label{results:technologies}

\begin{figure}[t]
\centering
\subfigure{
\includegraphics[width=.47\textwidth]{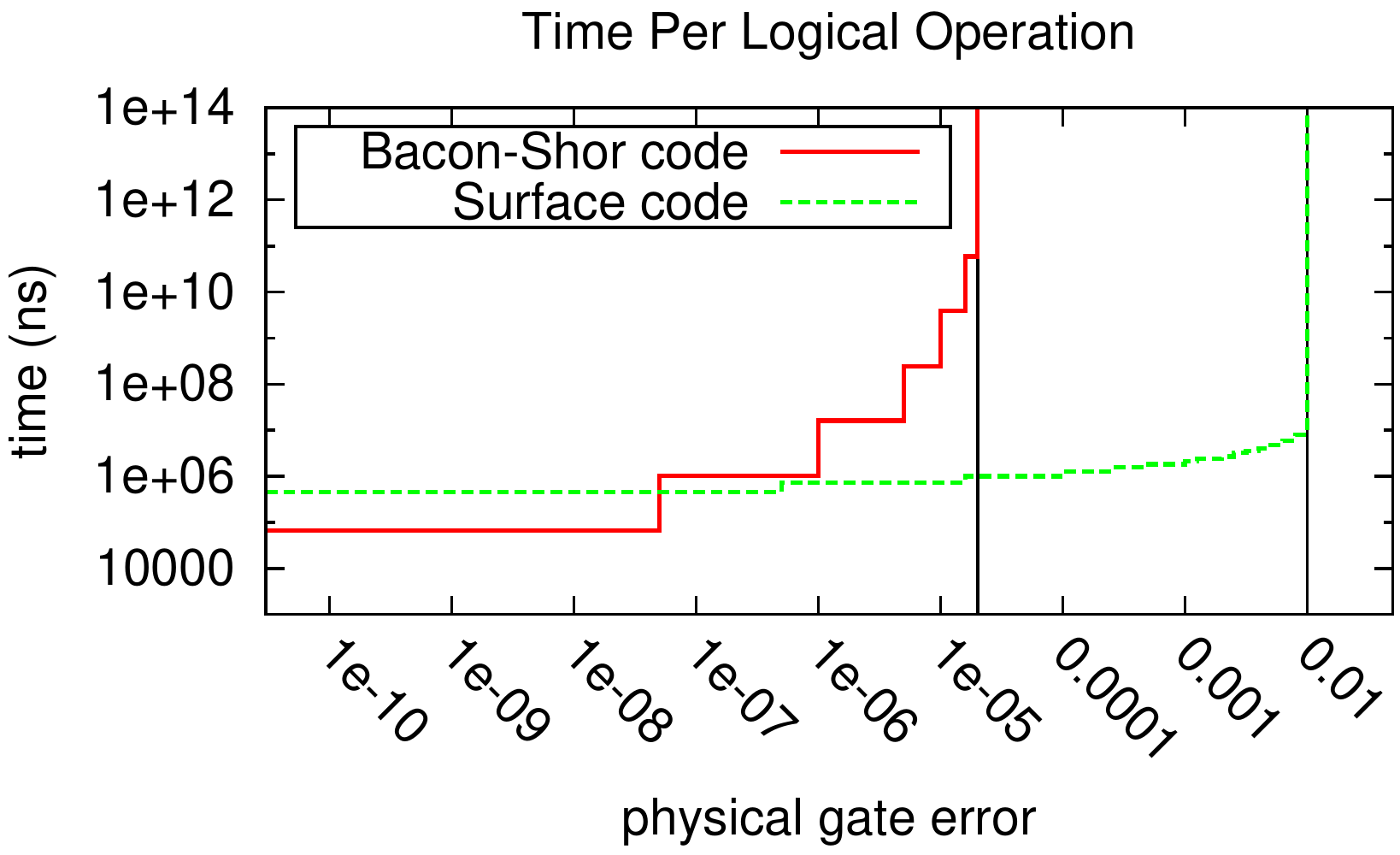}
}
\subfigure{
\includegraphics[width=.47\textwidth]{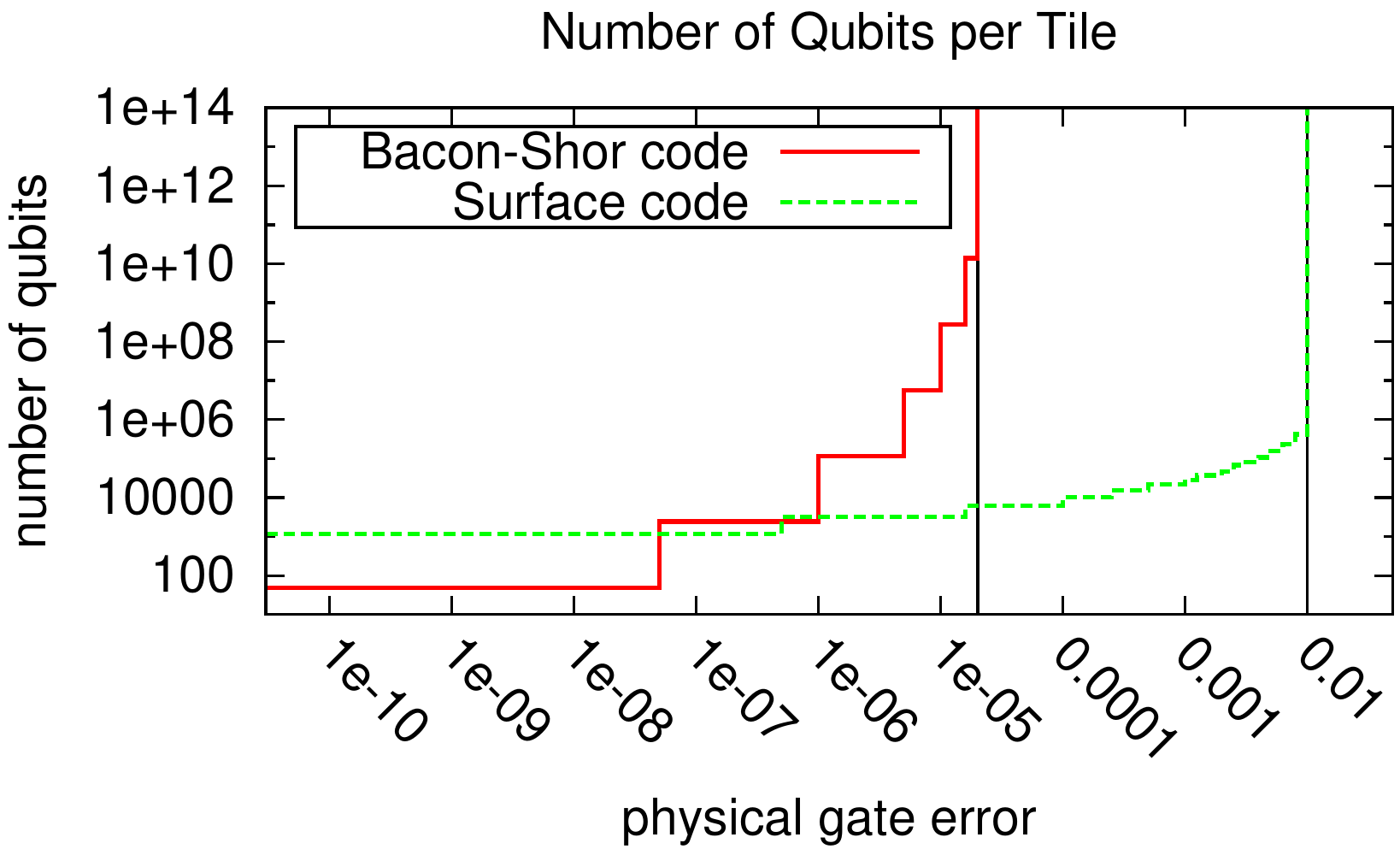}
}
\subfigure{
\includegraphics[width=.47\textwidth]{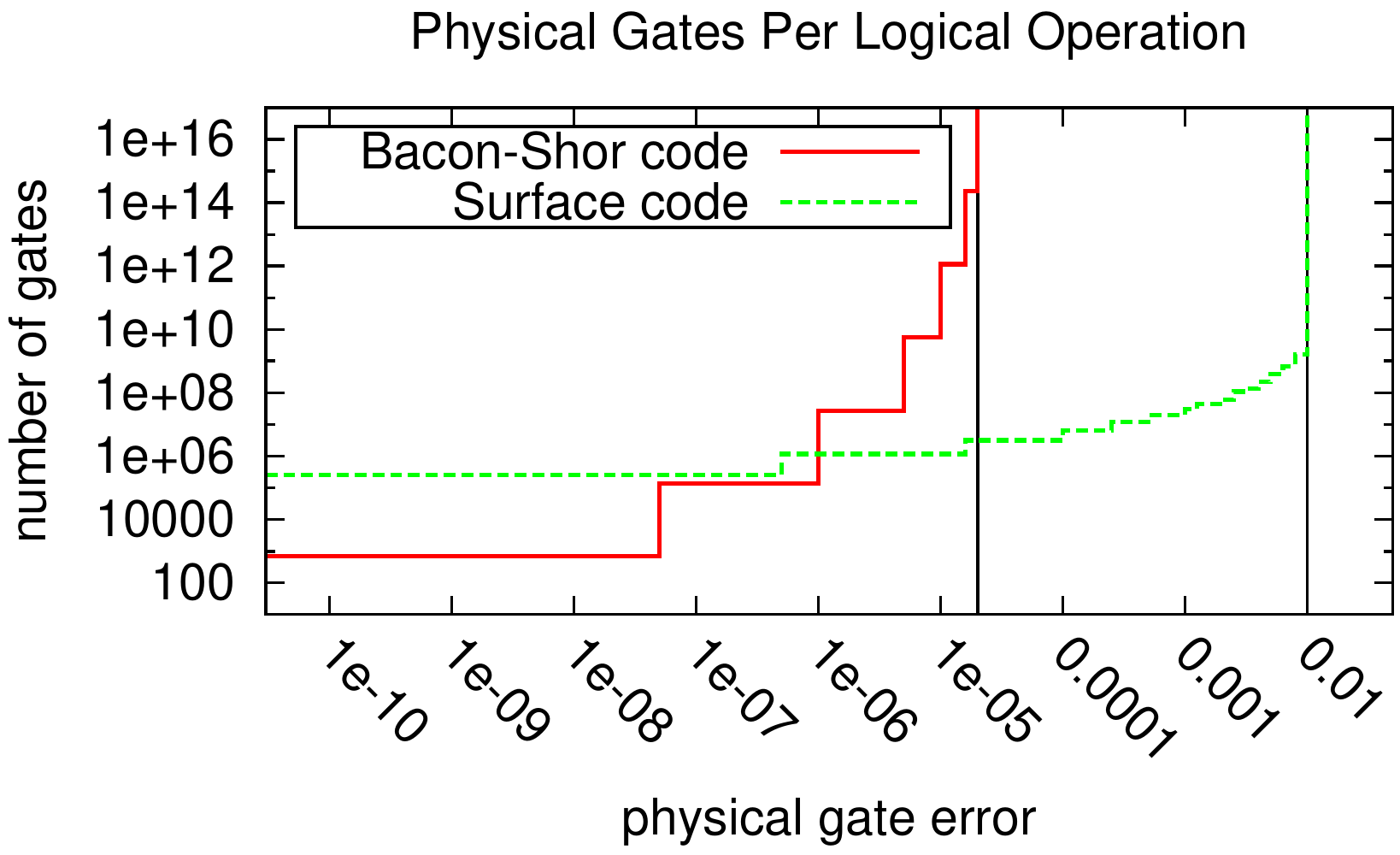}
}
\subfigure{
\includegraphics[width=.47\textwidth]{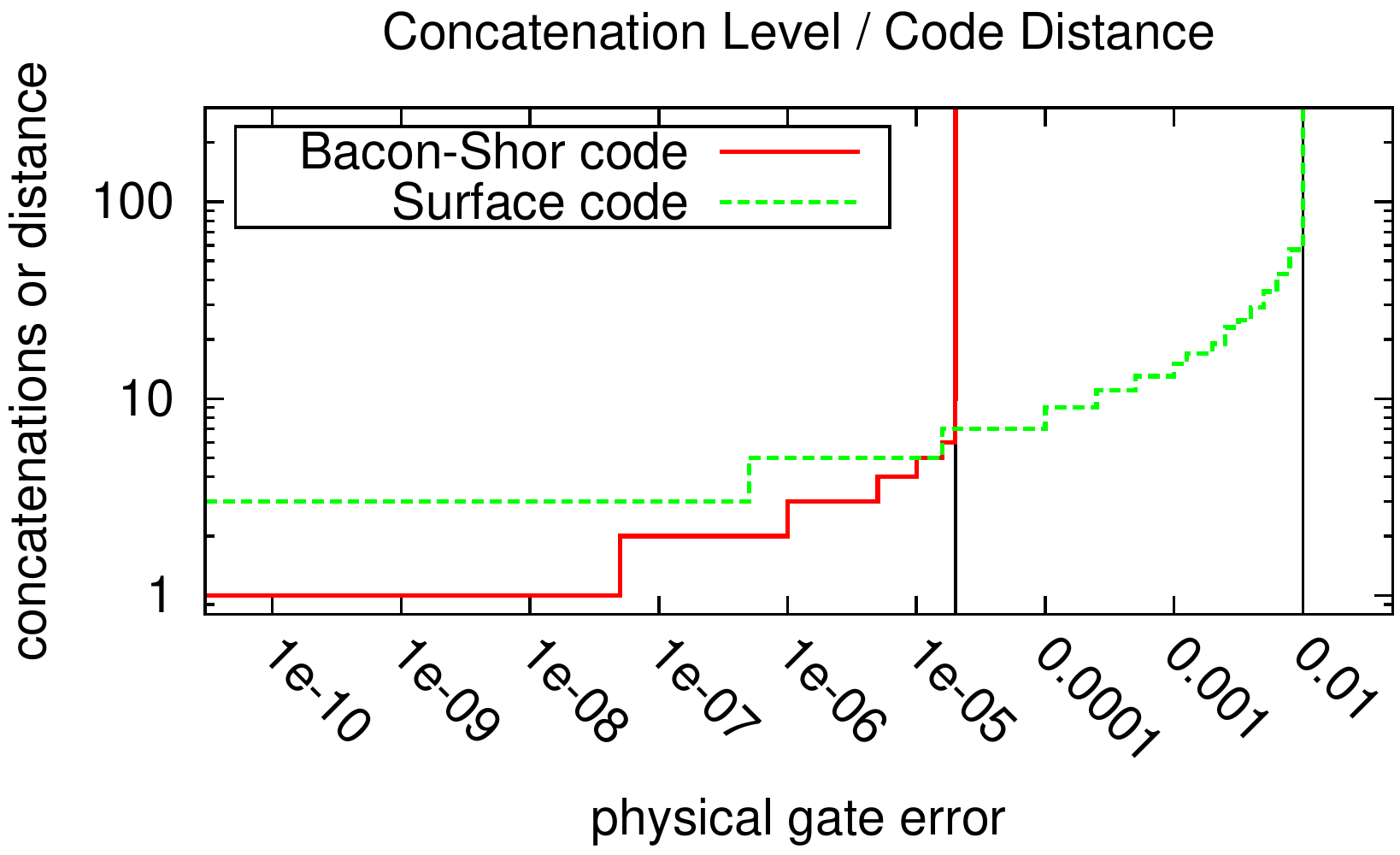}
}
\fcaption{Properties of error correction in an abstract quantum technology with physical gate error varying between $1 \times 10^{-10}$ and $1 \times 10^{-2}$. Vertical lines indicate the error correction threshold of the Bacon-Shor and Surface error-correcting codes.  The target error rate for a logical operation was chosen to be $1 \times 10^{-10}$.}
\label{fig:steps}
\end{figure}

Fig.~\ref{fig:steps} shows the crossover of the performance of the Bacon-Shor code and the surface code. The plots were obtained by simulating performance for an abstract quantum technology with gate error rates varying between $1 \times 10^{-10}$ and $1 \times 10^{-2}$ and physical gate durations fixed at $1,000$ ns. We also assume that the error correction by both codes needs to achieve a target error rate of $1 \times 10^{-10}$, which is sufficiently low to complete factoring successfully with high probability. The two vertical lines shown in the figure represent the thresholds of the Bacon-Shor and surface codes at $2.02 \times 10^{-5}$ and $1 \times 10^{-2}$, respectively. The codes are ineffective above the threshold. The shown time per logical operation and the number of physical gates per logical gate are averages over all logical gate types weighted by their occurrence in the most well-known quantum algorithms~\cite{analysis_BinWeldTree, analysis_BoolFormAlg, analysis_GroundStateEst, analysis_QuantLinSyst, analysis_ShortVecProb, analysis_QuantClassNum, analysis_TriangleFinding}. 

The crossover point of the time per logical operation, number of physical qubits per tile, and the number of physical gates per logical gate in Fig.~\ref{fig:steps} is around $1 \times 10^{-7}$. The surface code performs better for gate error rates above this value. The crossover occurs because the overhead of the concatenated Bacon-Shor code rises exponentially with the concatenation level, whereas it rises only moderately for surface codes. The last plot in Fig.~\ref{fig:steps} shows the concatenation level and code distance needed for each physical gate error rate, and it explains why the other metrics increase in steps.

\begin{figure}[h]
\centering
\subfigure{
\includegraphics[width=.47\textwidth]{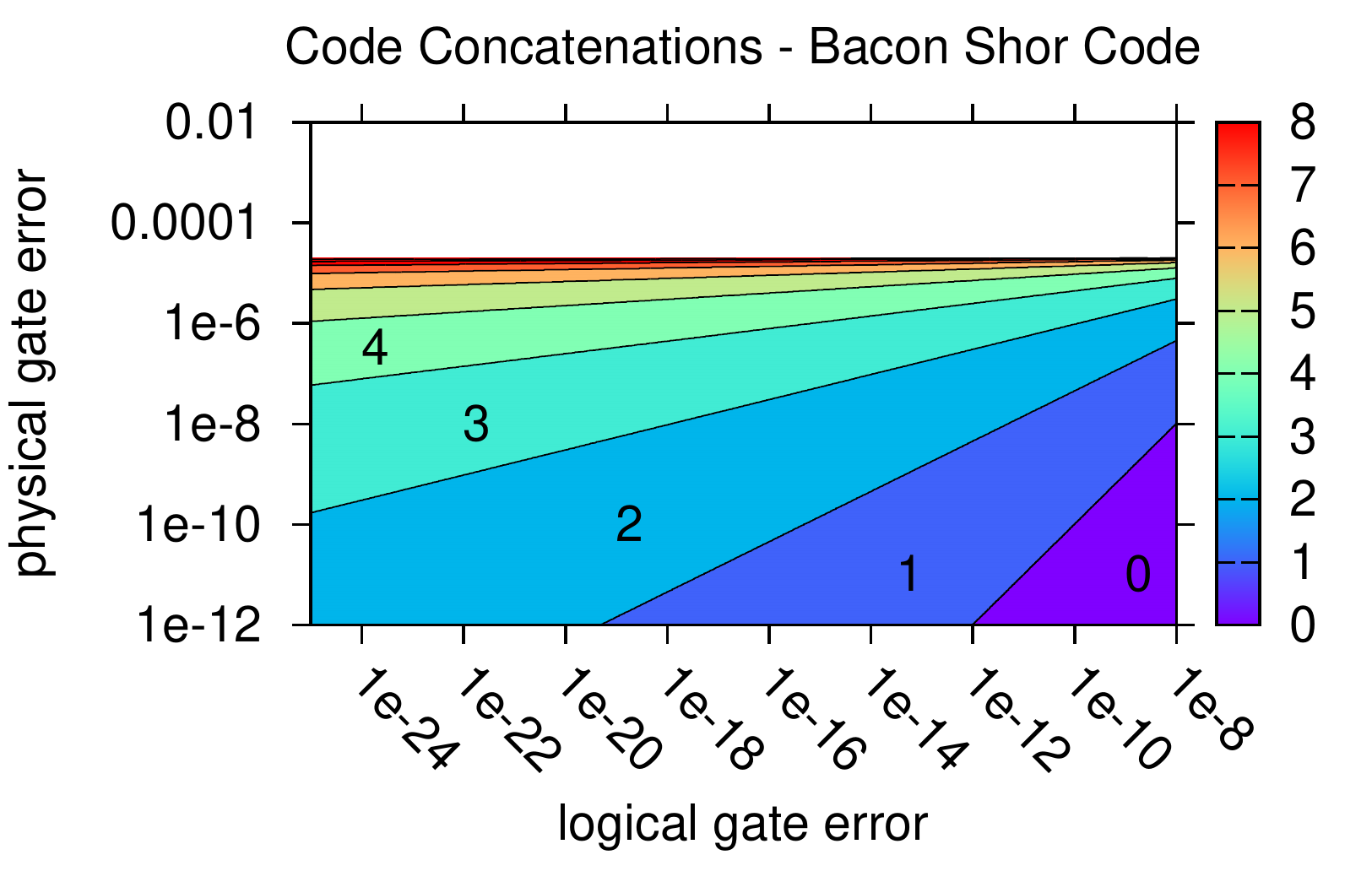}
}
\subfigure{
\includegraphics[width=.47\textwidth]{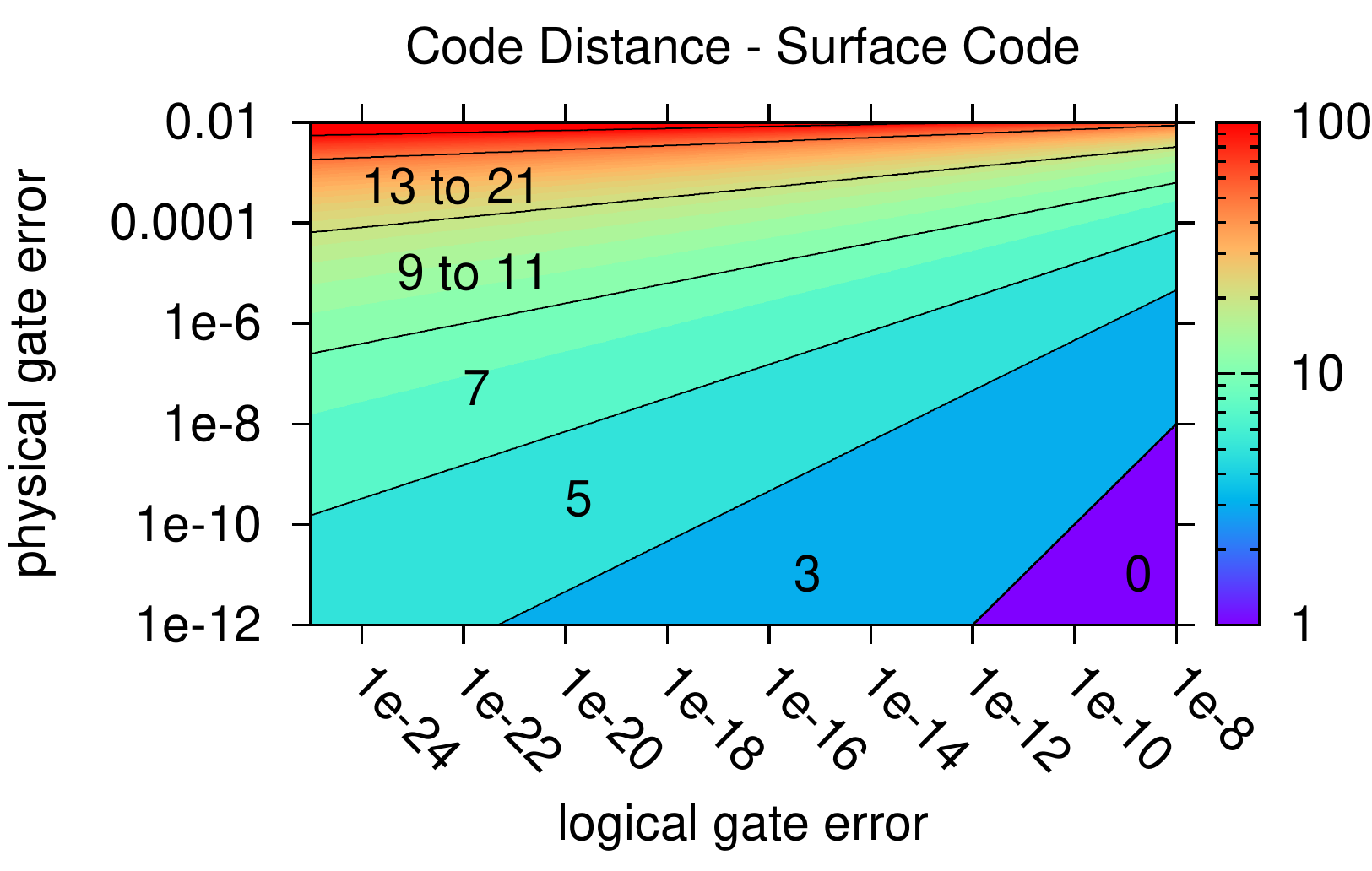}
}
\fcaption{The required concatenation level and code distance of the Bacon Shor and surface codes increase with increasing gate error of the physical technology and decreasing desired logical gate error.}
\label{fig:3DPlot}
\end{figure}

Seeing how profound an effect the concatenation level and code distance have on the resources, next we study how these values change as a function of varying target logical error rate as well as varying physical gate error rate. The results are shown in Fig.~\ref{fig:3DPlot}. When the physical gate error rate is lower or equal to the desired logical one, no error correction is needed. This is shown as $0$ levels of concatenation or $0$ distance. The concatenation level or code distance increases as  logical error rate decreases and physical error rate increases until the threshold of the quantum error-correcting code is exceeded.

\begin{figure}[h]
\centering
\subfigure{
\includegraphics[width=.47\textwidth]{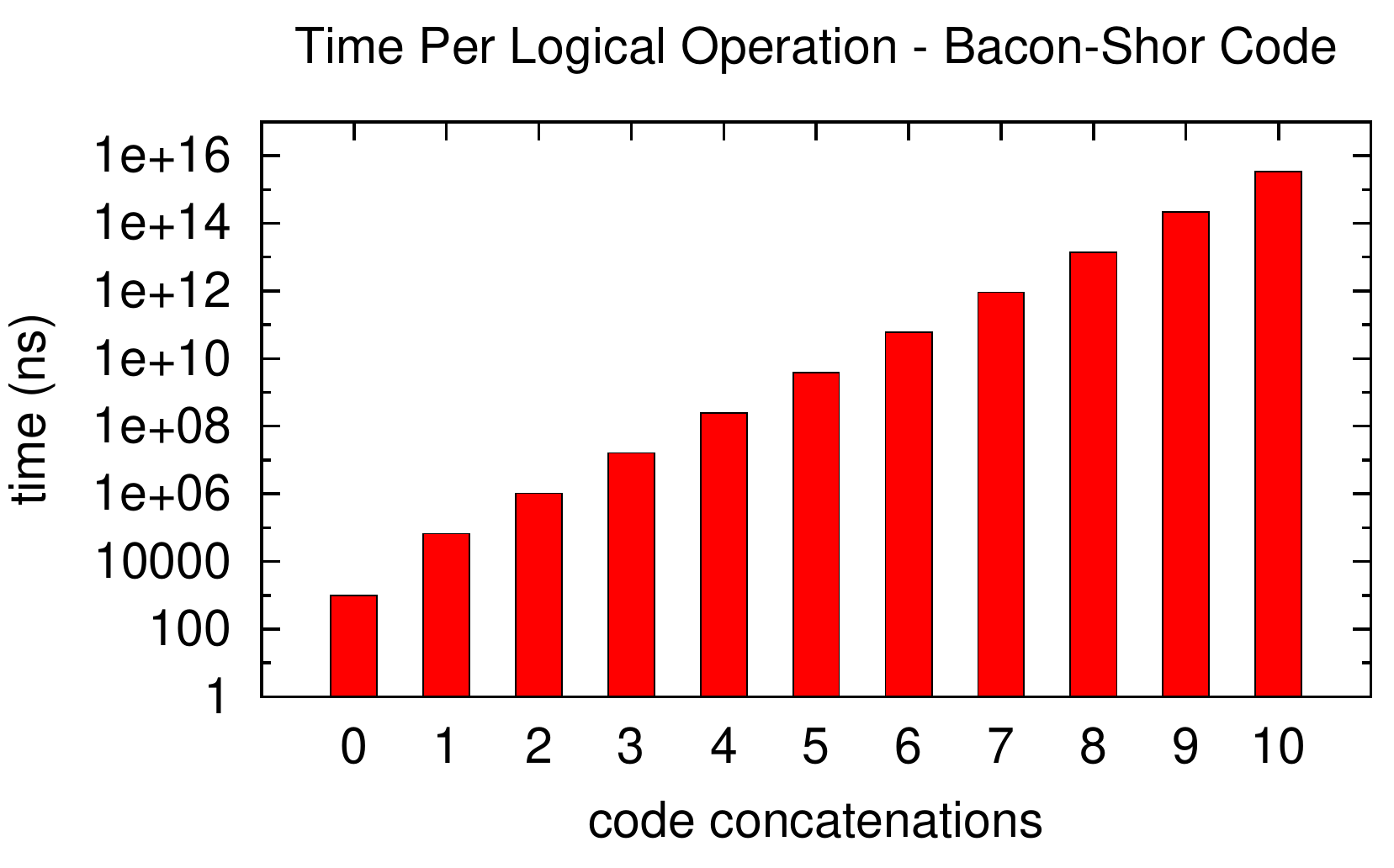}
}
\subfigure{
\includegraphics[width=.47\textwidth]{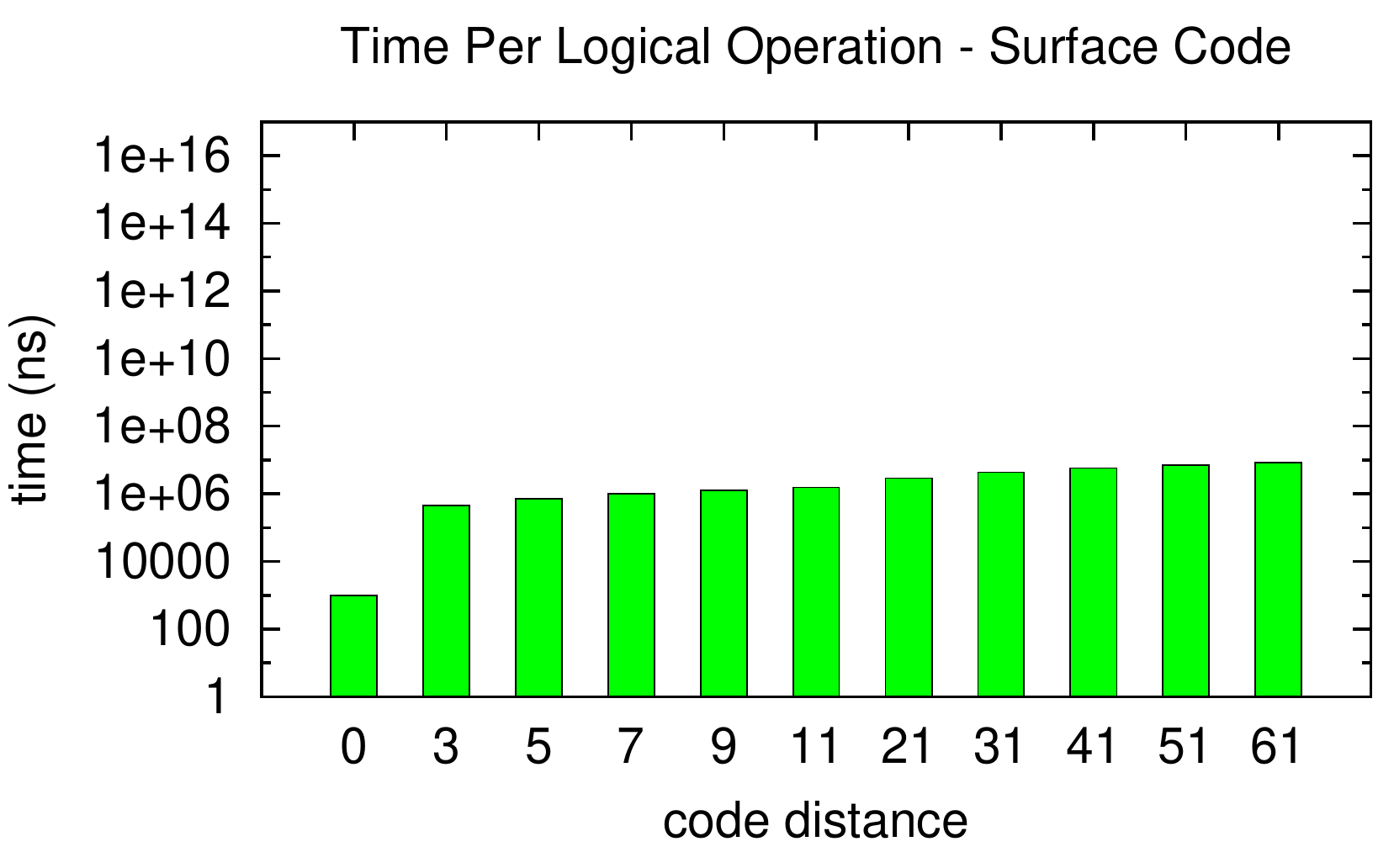}
}
\fcaption{The bar plots show the time needed for a logical operation at a given concatenation level or code distance.}
\label{fig:logicalGateTime}
\end{figure}

Fig.~\ref{fig:logicalGateTime} shows the duration of a logical operation with the Bacon-Shor and surface codes for a few code distances and concatenation levels. The values are averages over all logical gate types weighted by their occurrence in several representative quantum algorithms~\cite{analysis_BinWeldTree, analysis_BoolFormAlg, analysis_GroundStateEst, analysis_QuantLinSyst, analysis_ShortVecProb, analysis_QuantClassNum, analysis_TriangleFinding}. Concatenation level $0$ in Fig.~\ref{fig:logicalGateTime} means that no error correction is used, and the duration of the operation is equal to the time of the physical gate. The figure was obtained for a technology with physical gate time of $1,000$ ns. We observed that the logical gate time increases exponentially with increasing concatenation level for the Bacon-Shor code, but it only increases polynomially with increasing distance for the surface code.

\subsection{Comparison of Gate Composition}
\label{results:gates}

\begin{figure*}[t!]
\centering
\subfigure{
\includegraphics[width=.31\textwidth]{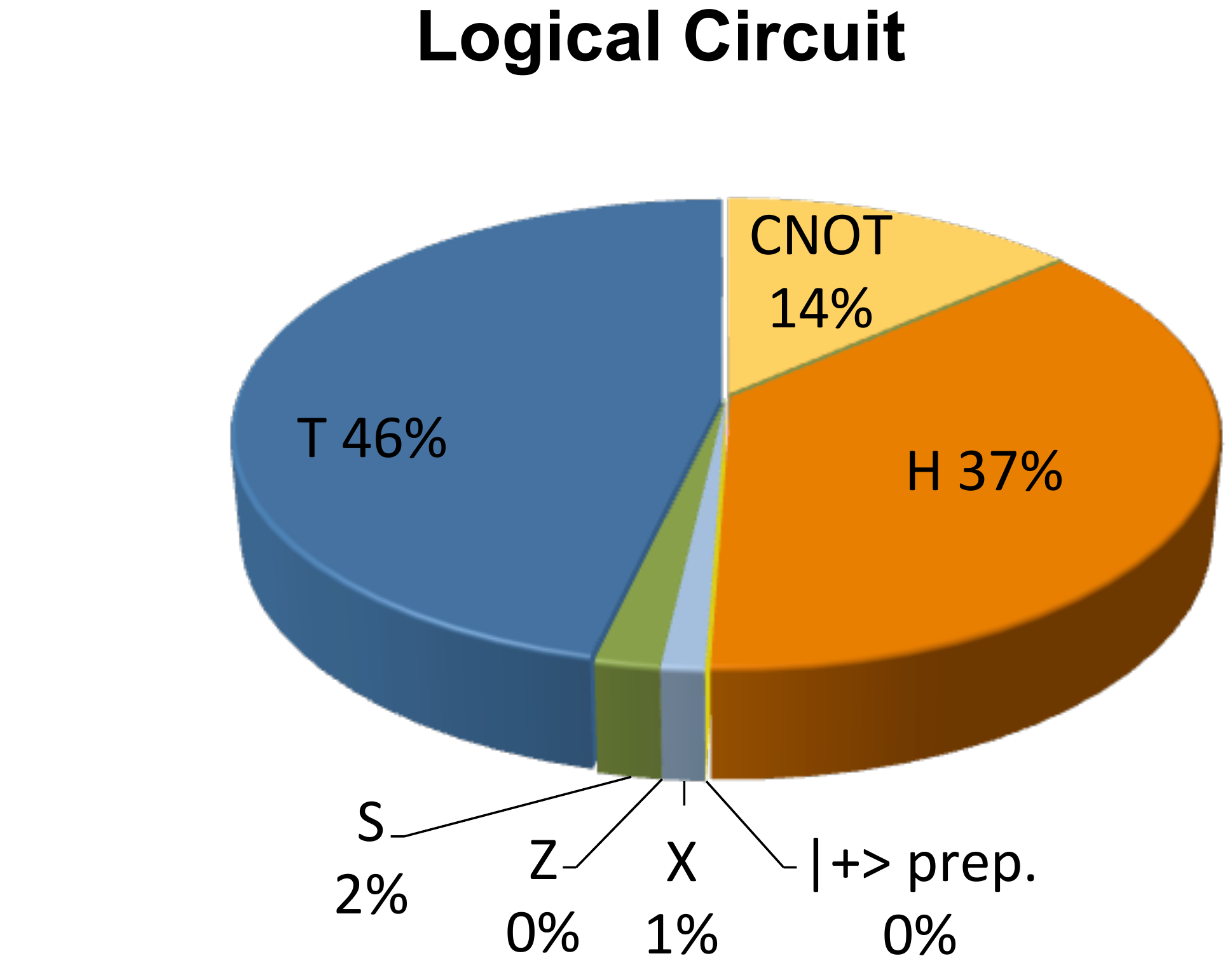}
}
\subfigure{
\includegraphics[width=.31\textwidth]{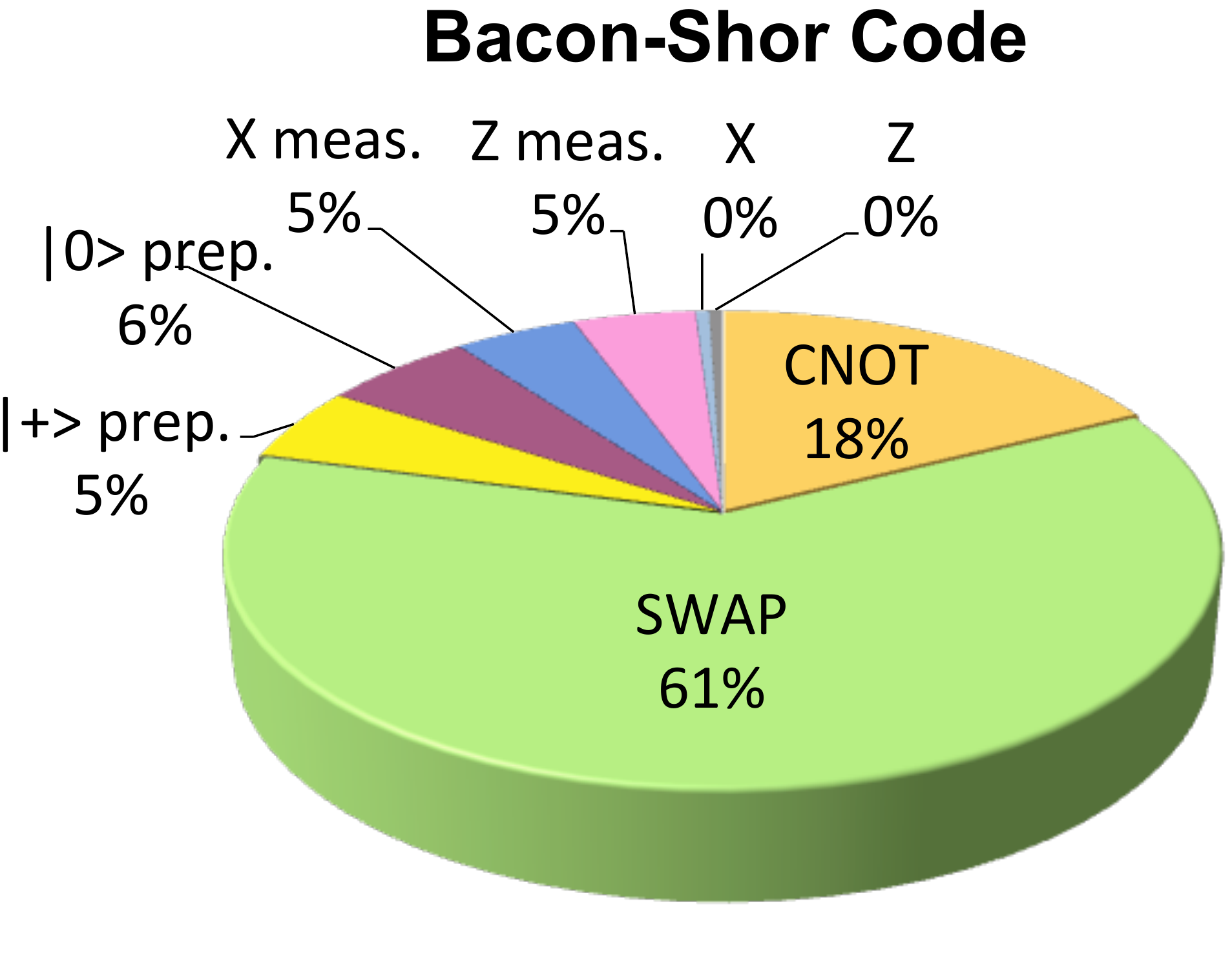}
}
\subfigure{
\includegraphics[width=.31\textwidth]{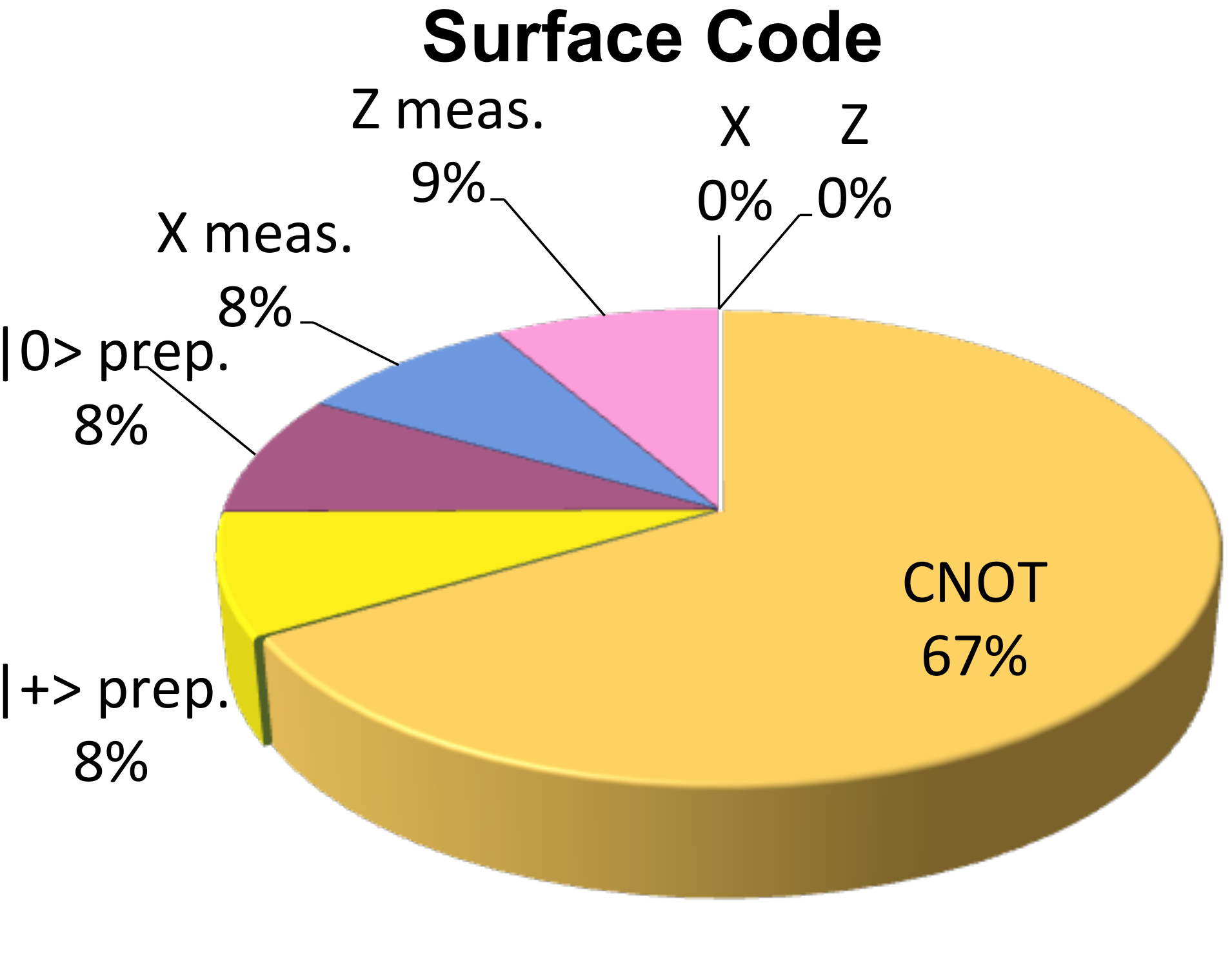}
}
\fcaption{The gate types used in a typical logical circuit and a typical fault-tolerant circuit that uses the Bacon-Shor and Surface codes all differ.}
\label{fig:pieCharts}
\end{figure*}

Fig.~\ref{fig:pieCharts} shows three pie charts with a breakdown of the gate types used by a circuit implementing a typical quantum algorithm, and fault-tolerant circuits that use the Bacon-Shor and surface codes. To capture the behavior in the typical case, the pie charts show average values for the algorithms studied in the QuRE toolbox~\cite{analysis_BinWeldTree, analysis_BoolFormAlg, analysis_GroundStateEst, analysis_QuantLinSyst, analysis_ShortVecProb, analysis_QuantClassNum, analysis_TriangleFinding}. The pie charts do not show the gate types that occurred with frequency lower than $0.01\%$.

Consistent with our expectation, $SWAP$ gates are frequently used by the Bacon-Shor code but not by the surface code. The Bacon-Shor code usess $SWAP$ gates to move qubits inside a tile before performing $CNOT$ gates. In contrast, the surface code is fully local and $CNOT$s are performed by braiding holes in the surface. We note that each SWAP gate can be replaced by three CNOT gates, avoiding the use of the SWAP. However, physical movement of qubits can lead to better performance in some physical technologies. We also observed in Fig.~\ref{fig:pieCharts} that the logical circuit contains a large fraction of $T$ gates and some $S$ gates, whereas the fraction of these gates in the fault-tolerant circuits is negligible. This is due to the fact that complex state distillation and one or more rounds of error correction are need for each logical $S$ or $T$ gate, and only a few applications of the $S$ and $T$ gate are needed. The frequently occurring error-correcting operation does not use any $S$ or $T$ gates.

\subsection{Some Limitations}
\label{results:limitations}

Our results obtained with the QuRE toolbox estimate the behavior of complex physical systems with millions of qubits and quantum gates and several choices of qualitatively different quantum technologies, algorithms, and quantum error-correcting codes. The QuRE toolbox must make a number of simplifying assumptions.

\textbf{Simplified logical qubit movement:} Because analyses of quantum algorithms typically do not specify layout of the logical qubits, we are unable to determine the average distance of logical qubits interacting in $CNOT$ gates. Because concatenated codes require that these interacting qubits be moved using logical $SWAP$ gates to neighboring locations, we may be underestimating the resources needed to do computation with concatenated codes.

\textbf{Simplified error model:} The toolbox assumes that all physical gates have an error rate equal to the error rate of the worst gate. Simulations use the depolarizing error model. These assumptions were made because the same assumptions were used in the literature that estimates thresholds of quantum error-correcting codes that we use. For technologies in which some gates are more reliable than others, it may be possible to use fewer concatenations or smaller code distance. Furthermore, optimizations can be made for cases with highly asymmetric error rates (e.g. when phase shift errors are much more likely than bit flips). Such optimizations for the Bacon-Shor code were done in~\cite{asymmetric_BS_1, asymmetric_BS_2}.

\textbf{Abstract models of quantum technologies:} The toolbox assumes that all quantum technologies can be fully described by specifying physical gate times and errors, and that qubits are arranged in a two-dimensional plane and neighboring qubits can always interact. In reality, the placement and movement of qubits is more restrictive, requiring additional resources to carry out two-qubit gates. A simple calculation that is specific to each particular technology allows us to estimate the additional overhead.

\textbf{Conservative estimate of circuit parallelism:} Resource reduction could be obtained by aggressive optimizations of the logical quantum circuits. In particular, we believe that many more logical gates can be performed in parallel. However, studying specific properties of any quantum algorithm is beyond the scope of this work.

\textbf{`Safe' choices for a range of parameters:} Our estimates use accepted methods such as performing error correction in case of concatenated codes after execution of every logical gate, and we also require that the quantum algorithm finishes successfully with high probability. Lower resource requirements could be achieved by relaxing these requirements. The distillation overhead of the surface code can be reduced by adjusting the code distance to the distillation level, as is suggested in~\cite{RHG:threshold}.

%% file: conclusion.tex
\section{Conclusion}
\label{conclusion}

In this article we analyzed the performance of concatenated and topological error-correcting codes. We found out that topological error correction works better for less reliable physical technologies, but concatenated codes are preferable for technologies with very low error rates. The decision which code to use depends on a number of parameters, and our work provides the tools necessary to perform simulations and systematically explore the design space. As the error rates of future quantum technologies are likely to improve, future research of error correction should focus on reducing the overhead of topological codes at low error rates. On the experimental side, we believe that fabricating one of two possible quantum technologies may be desirable. One of these technologies has unreliable quantum gates with error rates below the threshold of topological codes, but it has a very fast clock speed. Another good choice would be a reliable technology where slower clock speed is compensated by good support of qubit movement and extremely low error rates that allow efficient error correction with just a few concatenations of concatenated codes.

\nonumsection{Acknowledgements}
Development of the QuRE toolbox was supported by the IARPA QCS program (document numbers D11PC20165 and D11PC20167). The views and conclusions contained herein are those of the authors and should not be interpreted as necessarily representing the official policies or endorsements, either expressed or implied, of IARPA, DoI/NBC, or the U.S. Government. We appreciate the discussions with Sergey Bravyi, Kenneth Brown, Todd Brun, Austin Fowler, David Hocker, Daniel Lidar, Massoud Pedram, Robert Rausendorf and others who helped us to improve our methodology.